\documentclass{article}
\usepackage{amssymb,hyperref}
\usepackage[english]{babel}
\usepackage{amsmath, graphics}
\usepackage{geometry} 
\newcommand{\mathsym}[1]{{}}
\newcommand{\unicode}[1]{{}}

\def\noi{\noindent}
\def\nqq{\hspace{-2em}}

\def\barr{\left(\begin{array}}
\def\earr{\end{array}\right)}
\def\beq#1{\begin{equation}\label{#1}}
\def\eeq{\end{equation}}
\def\ber#1{\begin{eqnarray}\label{#1} &&\nqq}%   left alignment
\def\eer{\end{eqnarray}}

\newcommand{\bear}[1]{\begin{eqnarray}\label{#1}}
\newcommand{\ear}{\end{eqnarray}}

\catcode`\@=11 \@addtoreset{equation}{section}\catcode`\@=12
\newcommand{\N}{ {\mathbb N} }
\newcommand{\R}{ {\mathbb R} }

\newcommand{\fnm}{\footnotemark}
\newcommand{\fnt}{\footnotetext}

\begin{document}

 \vspace{15pt}

 \begin{center}
 \large\bf

 On generalized Melvin solution  for  the  Lie  algebra $E_6$

 \vspace{15pt}

 \normalsize\bf
        S. V. Bolokhov\fnm[1]\fnt[1]{bol-rgs@yandex.ru}$^{, b}$
        and   V. D. Ivashchuk\fnm[2]\fnt[1]{ivashchuk@mail.ru}$^{, a, b}$

 \vspace{7pt}

 \it (a) Center for Gravitation and Fundamental
 Metrology,  VNIIMS, 46 Ozyornaya St., Moscow 119361, Russia  \\

 (b) \ \ Peoples' Friendship University of Russia (RUDN University),
 6 Miklukho-Maklaya St.,  Moscow 117198, Russia \\

 \end{center}
 \vspace{15pt}

 \small\noi

 \begin{abstract}
  A multidimensional generalization of Melvin's
  solution for an arbitrary simple Lie algebra $\cal G$
  is considered. The gravitational model in $D$ dimensions, 
  $D \geq 4$, contains $n$ 2-forms and
 $l \geq n$ scalar fields, where $n$ is the rank of $\cal G$.
 The solution is governed by a set of $n$ functions $H_s(z)$ obeying
 $n$ ordinary differential equations with certain boundary conditions
 imposed. It was conjectured earlier that these functions should
 be polynomials (the so-called fluxbrane polynomials). 
 The polynomials $H_s(z)$, $s = 1,\dots,6$, for the Lie algebra $E_6$ are obtained
 and a corresponding solution for $l = n = 6$ is presented. 
 The polynomials depend upon integration constants $Q_s$, $s = 1,\dots,6$.
 They obey symmetry and duality identities. The latter ones are used 
 in deriving  asymptotic relations for solutions at large distances. 
  The power-law asymptotic relations for  $E_6$-polynomials at large $z$ 
  are governed by  integer-valued matrix  $\nu = A^{-1} (I + P)$,   where
  $A^{-1}$ is the inverse Cartan matrix,   $I$ is the identity matrix and
  $P$ is  permutation matrix, corresponding to a generator of the $Z_2$-group  
  of symmetry of the Dynkin diagram.   The 2-form fluxes $\Phi^s$, $s = 1,\dots,6$, are calculated.

 \end{abstract}

\large 

 \section{Introduction}

  In this paper we deal with a multidimensional generalization of    the Melvin solution \cite{Melv} which was considered   earlier in ref. \cite{GI-09}.   This  solution is governed by
  a simple finite-dimensional  Lie algebra. It is a special case of the so-called generalized fluxbrane solutions from \cite{Iflux}.  For generalizations of the Melvin solution,  fluxbrane solutions   and their applications, see  refs. 
  \cite{GW}-\cite{IM-fb-14} and the references therein. 

  We remind the reader that  Melvin's original solution in $4d$ space-time describes the gravitational field of a magnetic flux tube. The multidimensional analog of such a flux tube, supported by a certain configuration of 
  fields of forms, is  referred to as a fluxbrane (a ``thickened brane'' of magnetic flux). The appearance of fluxbrane solutions was motivated by superstring/M-theory models. A physical interest in such solutions is that they supply an appropriate background geometry for studying various processes 
  involving branes, instantons, Kaluza--Klein monopoles, pair production of magnetically charged  black holes and other configurations which can be studied via a special kind of Kaluza--Klein reduction (``modding technique'') of a certain multidimensional model in the presence of $U(1)$ isometry subgroup.
 
  The Melvin solution is geodesically complete \cite{MW}. Its group of isometry is $U(1)\times P(1,1)$,
  where $P(1,1)$ is $3$-dimensional isometry group of $2$-dimensional Minkowski space. $P(1,1)$   is semi-direct product of $O(1,1)$ and $\R^2$.

  In ref. \cite{GI-09} the electro-vacuum Melvin solution was generalized for the $D$-dimensional
  model which contains metric $g$, $n$  $2$-form fields $F^s = dA^s$ and $l$ scalar fields 
  $\varphi^{\alpha}$. The model also includes $n$   dilatonic coupling vectors belonging to $\R^l$. The  $D$-dimensional warped product solution from ref. \cite{GI-09}   comprises two factor spaces: $1$-dimensional subspace $M_1$ and a $(D-2)$-dimensional Ricci-flat subspace $M_2$.
  Here $M_1$ is either $\R$ or $S^1$. For $M_1 = S^1$ we have a cylindrically symmetric solution 
  with the isometry group $U(1)\times { \rm Isom}(M_2)$, where ${ \rm Isom}(M_2)$ is the isometry group of $M_2$.

  The generalized  fluxbrane solutions from ref. \cite{GI-09} are governed by functions
  $H_s(z) > 0$ defined on the interval $(0, +\infty)$ which obey the non-linear differential equations
  \beq{1.1}
  \frac{d}{dz} \left( \frac{ z}{H_s} \frac{d}{dz} H_s \right) =
   P_s \prod_{s' = 1}^{n}  H_{s'}^{- A_{s s'}},
  \eeq
 with  the following boundary conditions:
 \beq{1.2}
   H_{s}(+ 0) = 1,
 \eeq
 $s = 1,...,n$, where  $P_s > 0$  for all $s$.  Parameters  $P_s$ are proportional to
 $Q_s^2$, where $Q_s$ are integration constants  and $z = \rho^2$, where  
 $\rho$ is a radial parameter. The boundary condition
 (\ref{1.2}) guarantees the absence of a conic singularity (in the metric) for $\rho =  +0$.
The integration constants $Q_s$ are coinciding up to a sign with values of  magnetic fields 
on the axis of the symmetry.

 In this paper we assume that $(A_{s s'})$ is a Cartan matrix for some 
 simple finite-dimensional Lie algebra $\cal G$ of rank $n$ ($A_{ss} = 2$ for all $s$).
 
 According to a conjecture  suggested in \cite{Iflux}, the
 solutions to Eqs. (\ref{1.1}), (\ref{1.2}) governed by the Cartan matrix $(A_{s s'})$
 are  polynomials: 
  \beq{1.3}
  H_{s}(z) = 1 + \sum_{k = 1}^{n_s} P_s^{(k)} z^k,
  \eeq
  where $P_s^{(k)}$ are constants ($P_s^{(1)} = P_s$). Here
 $P_s^{(n_s)} \neq 0$  and 
 \beq{1.4}
 n_s = 2 \sum_{s' =1}^{n} A^{s s'}
 \eeq 
 where we denote $(A^{s s'}) = (A_{s s'})^{-1}$.
 Integers $n_s$ are components  of a twice dual
 Weyl vector in the basis of simple co-roots \cite{FS}.
 
The set of fluxbrane polynomials $H_s$ defines a 
special solution to  open Toda chain equations \cite{K,OP} corresponding 
to a simple finite-dimensional Lie algebra $\cal G$; see ref. \cite{I-14}.
In refs. \cite{GI-09,GI} a program  (in Maple)  for calculation of these polynomials for
classical series of Lie  algebras (of $A$-, $B$-, $C$- and $D$-series) was suggested.

It should be noted that  the open Toda chain corresponding  to the Lie algebra $\cal G$ 
has a hidden symmetry group $G_{T} = \exp(\cal G)$. The solution from ref. \cite{GI-09}
corresponding to this group is a special case of solutions from \cite{Iflux}. It  may be obtained by using an $1$-dimensional sigma-model  \cite{IMC,IMJ,IK} with $(2 + l + n)$-dimensional target space.
 The isometry group of this target space $G_{sm}$ (related to the sigma model) was studied in detail in \cite{I-98}. 
For another more general setup with non-diagonal metrics (which is valid for flat $M_2$)  see also \cite{GR}.
The group $G_{sm}$  is another hidden symmetry group related to our model.
Here the Toda  Lagrangian $L_T$ may be obtained from the sigma-model one after
integrating the Maxwell-type equations corresponding to potentials $\Phi^s(u) = A^s_{\phi}(u)$,
where $u$ is a radial variable and $\phi$ is a coordinate on $M_1$ ($0 < \phi < 2 \pi$ for $M_1 = S^1$),
and obtaining integration constants $Q_s$. The Toda  Lagrangian $L_T = L_T(x,\dot{x},Q)$ ($\dot{x} = \frac{dx}{du}$)
 is responsible for equations of motion for $2$ scale factors and $l$ scalar fields described by $x = (x^a)$
for fixed $Q = (Q_s)$.  

We note also that there are several multidimensional aspects of generalized Melvin solution from ref.  \cite{GI-09}: 
(1) the space-time dimension  $D$ (for Melvin's solution $D = 4$), 
(2) the rank of the Toda group $G_{T}$ which is equal to $n$ (in Melvin's case $n = 1$) and 
(3) the dimension of the target space of the corresponding  sigma-model which is equal to $N = n + l + 2$ 
(in Melvin's case $N = 3$). 
 
  Here we verify the conjecture from ref. \cite{Iflux} for the Lie algebra $E_6$.
  In Section 2 the generalized  Melvin 
  solution for an arbitrary simple finite-dimensional Lie algebra $\cal G$
  is considered.  The exact solution for the  Lie algebra $E_6$ is presented
  in Section 3, while the fluxbrane polynomials are listed in the Appendix.
  Here  duality relations for the polynomials $H_s(z)$ and asymptotic formulas
  for $z \to + \infty$ are presented, as well as the asymptotics for the solutions
  at large distances and a calculation of flux integrals. We  find that   
  any  flux $\Phi^s$ depends   upon  the integration constant  $Q_s$ and does not
  depend upon the other constants  $Q_{s'}$, $s' \neq s$. The flux $\Phi^s$ is proportional to $n_s Q_s^{-1}$,
  where $n_s$ are integer numbers (\ref{1.4}):  $n_s =16,30,42,30,16,22$ for $s =1,2,3,4,5,6$, respectively.

\section{The main solution}

We consider a  model governed by the action
 \beq{2.1}
  S=\int d^Dx \sqrt{|g|} \biggl \{R[g]-
  h_{\alpha\beta}g^{MN}\partial_M\varphi^{\alpha}\partial_N\varphi^{\beta}-\frac{1}{2}
  \sum_{s =1}^{n}\exp[2\lambda_s(\varphi)](F^s)^2 \biggr \},
 \eeq
where $g=g_{MN}(x)dx^M\otimes dx^N$ is a metric,
 $\varphi=(\varphi^\alpha)\in\R^l$ is a vector of scalar fields,
 $(h_{\alpha\beta})$ is a  constant symmetric non-degenerate
 $l\times l$ matrix $(l\in \N)$,    $ F^s =    dA^s
          =  \frac{1}{2} F^s_{M N}  dx^{M} \wedge  dx^{N}$
 is a $2$-form,  $\lambda_s$ is a 1-form on $\R^l$:
 $\lambda_s(\varphi)=\lambda_{s \alpha}\varphi^\alpha$,
 $s = 1,\dots, n$; $\alpha=1,\dots,l$.
 In (\ref{2.1}),
 we denote $|g| =   |\det (g_{MN})|$, $(F^s)^2  =
  F^s_{M_1 M_{2}} F^s_{N_1 N_{2}}  g^{M_1 N_1} g^{M_{2} N_{2}}$, $s = 1,\dots, n$.

 Here we consider a family of exact
solutions to the field equations corresponding to the action
(\ref{2.1}) and depending on one variable $\rho$. The solutions
are defined on the manifold
 \beq{2.2}
  M = (0, + \infty)  \times M_1 \times M_2,
 \eeq
 where $M_1$ is a one-dimensional manifold (say $S^1$ or $\R$) and
 $M_2$ is a (D-2)-dimensional Ricci-flat manifold. The solution
 reads \cite{GI-09}
 \bear{2.30}
  g= \Bigl(\prod_{s = 1}^{n} H_s^{2 h_s /(D-2)} \Bigr)
  \biggl\{ w d\rho \otimes d \rho  +
  \Bigl(\prod_{s = 1}^{n} H_s^{-2 h_s} \Bigr) \rho^2 d\phi \otimes d\phi +
    g^2  \biggr\},
 \\  \label{2.31}
  \exp(\varphi^\alpha)=
  \prod_{s = 1}^{n} H_s^{h_s  \lambda_{s}^\alpha},
 \\  \label{2.32a}
  F^s= - Q_s \left( \prod_{s' = 1}^{n}  H_{s'}^{- A_{s
  s'}} \right) \rho d\rho \wedge d \phi,
  \ear
  $\alpha = 1, \dots, l$ and $s = 1,\dots, n$, where $w = \pm 1$, $g^1 = d\phi \otimes d\phi$ is a
  metric on $M_1$ and $g^2$ is a  Ricci-flat metric on
 $M_{2}$.

 The functions $H_s(z) > 0$, $z = \rho^2$, obey the equations
(\ref{1.1}) with the boundary conditions (\ref{1.2}) and
 \beq{2.21}
  P_s =  \frac{1}{4} K_s Q_s^2.
 \eeq
 The parameters  $h_s$  satisfy the relations
  \beq{2.16}
  h_s = K_s^{-1}, \qquad  K_s = B_{s s} > 0,
  \eeq
 where
 \beq{2.17}
  B_{ss'} \equiv
  1 +\frac{1}{2-D}+  \lambda_{s \alpha} \lambda_{s' \beta}   h^{\alpha\beta},
  \eeq
 $s, s' = 1,\dots, n$, with $(h^{\alpha\beta})=(h_{\alpha\beta})^{-1}$.
 Here
 $\lambda_{s}^{\alpha} = h^{\alpha\beta}  \lambda_{s \beta}$
 and
 \beq{2.18}
  (A_{ss'}) = \left( 2 B_{s s'}/B_{s' s'} \right)
 \eeq
  is the Cartan matrix for a simple Lie algebra $\cal G$ of rank $n$.

It may be shown that if the matrix $(h_{\alpha\beta})$
 has an Euclidean signature and  $l \geq n$, there exists a set of co-vectors
 $\lambda_1,\dots , \lambda_n$  obeying (\ref{2.18}).
 Thus the solution is valid at least when $l \geq n$
 and the matrix $(h_{\alpha\beta})$ is  positive-definite.

The solution under consideration is as a special case of the
fluxbrane (for $w = +1$,  $M_1 = S^1$) and $S$-brane
($w = -1$) solutions from \cite{Iflux} and \cite{GIM}, respectively.

If $w = +1$ and the (Ricci-flat) metric $g^2$ has a
pseudo-Euclidean signature, we get a multidimensional 
generalization of  Melvin's solution \cite{Melv}. 
 
Melvin's solution (without scalar field) corresponds to $D = 4$,
$n = 1$, $M_1 = S^1$ ($0 < \phi <  2 \pi$),  $M_2 = \R^2$,
$g^2 = -  dt \otimes dt + d \xi \otimes d \xi$
and ${\cal G} = A_1$. 
  
For $w = -1$ and $g^2$ of Euclidean signature we obtain a cosmological solution
with a horizon (as $\rho = + 0$) if $M_1 = \R$ ($ - \infty < \phi < + \infty$).

\section{The solution  for  the Lie algebra $E_6$}

Here we deal with the solution for $n = l = 6$, $w = + 1$ and $M_1 = S^1$,  which corresponds to the Lie algebra 
$E_6$. We put here $h_{\alpha\beta} = \delta_{\alpha \beta}$ and denote  
$(\lambda_{s a }) =  (\lambda_{s}^{a}) = \vec{\lambda}_{s}$, $s = 1, \dots, 6$.

  The matrix  $A=  (A_{ss'})$ is coincides with the Cartan matrix 
  for the exceptional Lie algebra $E_6$ 	
  
   \begin{equation}
 \label{3.1}
 A=  (A_{ss'}) = \left(
 \begin{array}{cccccc}
  2 & -1 & 0 & 0 & 0 & 0 \\
  -1 & 2 & -1 & 0 & 0 & 0 \\
  0 & -1 & 2 & -1 & 0 & -1 \\
  0 & 0 & -1 & 2 & -1 & 0 \\
  0 & 0 & 0 & -1 & 2 & 0 \\
  0 & 0 & -1 & 0 & 0 & 2
 \end{array}
 \right).
 \end{equation}
 
 This matrix  is graphically depicted at Fig. 1  by the Dynkin diagram.
  
  \vspace{60pt}
  
  \setlength{\unitlength}{1mm}
 \begin{figure}[h]
 \centering
 \begin{picture}(40, 0)
 \put(1,5){\circle*{2}}
 \put(11,5){\circle*{2}}
 \put(21,5){\circle*{2}}
 \put(31,5){\circle*{2}}
 \put(41,5){\circle*{2}}
 \put(21,15){\circle*{2}}
 \put(0,1){$1$}
 \put(10,1){$2$}
 \put(20,1){$3$}
 \put(30,1){$4$}
 \put(40,1){$5$}
 \put(20,17){$6$}
 \put(1,5){\line(10,00){40}}
 \put(20.9,5){\line(0,10){10}}
 \end{picture}
 \caption{The Dynkin diagram for the Lie algebra $E_6$.}
 \end{figure}
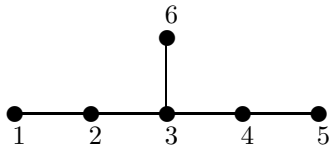
  
  \vspace{10pt}
 
 \subsection{Fluxbrane polynomials  for Lie algebra $E_6$}
 
  The inverse Cartan matrix for $E_6$
 \begin{equation}
 \label{3.2}
  A^{-1}= (A^{ss'}) = %\left(
 %\begin{array}{cccccc}
  \begin{pmatrix} \vspace{0.3em}
  \frac{4}{3} & \frac{5}{3} & 2 & \frac{4}{3} & \frac{2}{3} & 1 \\ \vspace{0.3em}
  \frac{5}{3} & \frac{10}{3} & 4 & \frac{8}{3} & \frac{4}{3} & 2 \\ \vspace{0.3em}
  2 & 4 & 6 & 4 & 2 & 3 \\ \vspace{0.3em}
  \frac{4}{3} & \frac{8}{3} & 4 & \frac{10}{3} & \frac{5}{3} & 2 \\ \vspace{0.3em}
  \frac{2}{3} & \frac{4}{3} & 2 & \frac{5}{3} & \frac{4}{3} & 1 \\ \vspace{0.3em}
  1 & 2 & 3 & 2 & 1 & 2
  \end{pmatrix}
 %\end{array}
 %\right)
 \end{equation}
  
 implies due to (\ref{1.4}) 
 \begin{equation}
  \label{3.3}
 (n_1,n_2,n_3,n_4,n_5,n_6) =  (16,30,42,30,16,22).
 \end{equation}
  
 For the Lie algebra $E_6$ we find the  set of six fluxbrane polynomials, which are
 listed in the appendix. Here as in \cite{I-14} we  parametrize the polynomials by using
 other parameters (here denoted $B_s$) instead of $P_s$:
 \begin{equation}
   \label{3.3a}
  P_s = n_s B_s,
  \end{equation}
 $s = 1, \dots, 6$.  This is necessary to avoid huge denominators in monomials of $H_s$.

  The polynomials have the following structure:
\begin{multline*}
H_1 = 1 +
16 B_1 z+ 120 B_1 B_2 z^2+ \dots +120 B_1^2 B_2^3 B_3^4 B_4^2 B_5 B_6^2 z^{14}\\
+16 B_1^2 B_2^3 B_3^4 B_4^3 B_5 B_6^2 z^{15}+B_1^2 B_2^3 B_3^4 B_4^3 B_5^2 B_6^2 z^{16},
%\label{3.4a1}
\end{multline*}\vspace{-2.5em}
\begin{multline*}
H_2 = 1 +
30 B_2 z+(120 B_1 B_2+315 B_3 B_2) z^2 \\ \dots +(120 B_1^3 B_2^6 B_4^5 B_5^2
B_6^4 B_3^8 +315 B_1^3 B_2^6 B_4^5 B_5^3 B_6^4 B_3^7) z^{28} \\
+ 30 B_1^3 B_2^6 B_3^8 B_4^5 B_5^3 B_6^4 z^{29}+B_1^3 B_2^6 B_3^8 B_4^6 B_5^3 B_6^4 z^{30},
%\label{3.4a2}
\end{multline*}\vspace{-2.5em}
\begin{multline*}
H_3 = 1 + 42 B_3 z+(315 B_2 B_3+315 B_4 B_3+231 B_6 B_3) z^2 \\ \dots +
(315 B_1^4 B_2^7 B_4^8 B_5^4 B_6^6 B_3^{11}
+315 B_1^4 B_2^8 B_4^7 B_5^4 B_6^6 B_3^{11}+231 B_1^4 B_2^8 B_4^8 B_5^4 B_6^5 B_3^{11}) z^{40} \\
+42 B_1^4 B_2^8 B_3^{11} B_4^8 B_5^4 B_6^6 z^{41}
+B_1^4 B_2^8 B_3^{12} B_4^8 B_5^4 B_6^6 z^{42},
 %\label{3.4a3}
\end{multline*}\vspace{-2.5em}
\begin{multline*}
H_4 = 1+
30 B_4 z+(315 B_3 B_4+120 B_5 B_4) z^2 \\ \dots +(120 B_1^2 B_2^5 B_4^6 B_5^3
B_6^4 B_3^8+315 B_1^3 B_2^5 B_4^6 B_5^3 B_6^4 B_3^7) z^{28} \\
+30 B_1^3 B_2^5 B_3^8 B_4^6 B_5^3 B_6^4 z^{29}+B_1^3 B_2^6 B_3^8 B_4^6 B_5^3 B_6^4 z^{30},
 %\label{3.4a4}
\end{multline*}\vspace{-2.5em}
\begin{multline*}
H_5 = 1 +
16 B_5 z+120 B_4 B_5 z^2+ \dots +120 B_1 B_2^2 B_3^4 B_4^3 B_5^2 B_6^2z^{14}\\
+16 B_1 B_2^3 B_3^4 B_4^3 B_5^2 B_6^2 z^{15}+B_1^2 B_2^3 B_3^4 B_4^3 B_5^2 B_6^2 z^{16},
 %\label{3.4a5}
\end{multline*}\vspace{-2.5em}
\begin{multline}
H_6 = 1 +
 22 B_6 z+231 B_3 B_6 z^2+ \dots + 231 B_1^2 B_2^4 B_3^5 B_4^4 B_5^2
 B_6^3 z^{20}\\
 +22 B_1^2 B_2^4 B_3^6 B_4^4 B_5^2 B_6^3 z^{21}+B_1^2 B_2^4 B_3^6 B_4^4 B_5^2 B_6^4 z^{22}.
\label{3.4}
\end{multline}

The powers of polynomials are in agreement with the relation (\ref{3.3}).
In what follows we denote
\begin{equation}
  \label{3.5}
   H_s = H_s(z) = H_s(z, (B_i) ),
 \end{equation}
$s = 1, \dots, 6$; where $(B_i) = (B_1,B_2,B_3,B_4,B_5,B_6)$.

Due to  (\ref{3.4}) the polynomials have the following asymptotical behavior 
\begin{equation}
  \label{3.6}
   H_s = H_s(z, (B_i) )  \sim \left( \prod_{l=1}^{6} (B_l)^{\nu^{sl}} \right) z^{n_s} \equiv 
   H_s^{as}(z, (B_i)),
 \end{equation}
$s = 1, \dots, 6$, as $z \to \infty$. Here
\begin{equation}
   \label{3.7}
   \nu = (\nu^{sl})  = \left(
   \begin{array}{cccccc}
    2 & 3 & 4 & 3 & 2 & 2 \\
    3 & 6 & 8 & 6 & 3 & 4 \\
    4 & 8 & 12& 8 & 4 & 6 \\
    3 & 6 & 8 & 6 & 3 & 4 \\
    2 & 3 & 4 & 3 & 2 & 2 \\
    2 & 4 & 6 & 4 & 2 & 4
   \end{array}
   \right).
   \end{equation}

The matrix (\ref{3.7}) is related to the inverse Cartan matrix as follows:
\begin{equation}
  \label{3.8}
  \nu = A^{-1} (I + P),
 \end{equation}
 where $I$ is $6 \times 6$ identity matrix and
\begin{equation}
   \label{3.9}
   P  = \left(
   \begin{array}{cccccc}
    0 & 0 & 0 & 0 & 1 & 0 \\
    0 & 0 & 0 & 1 & 0 & 0 \\
    0 & 0 & 1 & 0 & 0 & 0 \\
    0 & 1 & 0 & 0 & 0 & 0 \\
    1 & 0 & 0 & 0 & 0 & 0 \\
    0 & 0 & 0 & 0 & 0 & 1
   \end{array}
   \right).
   \end{equation}
is permutation matrix. This matrix corresponds to the permutation 
$\sigma \in S_6$ ($S_6$ is the symmetric group)
\begin{equation}
  \label{3.10}
   \sigma: (1,2,3,4,5,6) \mapsto (5,4,3,2,1,6),
 \end{equation}
 by the relation  $P = (P^i_j) = (\delta^i_{\sigma(j)})$. Here $\sigma$   is the generator of the group 
 $G = \{ \sigma, id \}$ which is the symmetry group of the Dynkin diagram. $G$ is isomorphic to the group $Z_2$.
 $\sigma$ is a composition of two transpositions: $(1 \ 5)$ and $(2 \ 4)$.
 
 We note that the matrix $\nu$ is symmetric  and 
 \begin{equation}
   \label{3.10a}
    \sum_{s= 1}^6 \nu^{sl} = n_l,
  \end{equation}
  $l = 1, \dots, 6$.
  
 Let us denote $\hat{B}_i = B_{\sigma(i)} $, $i= 1, \dots, 6$.
 We call the ordered set $(\hat{B}_i)$ a dual one to the ordered set $(B_i)$.
 By using the relations for  polynomials from the appendix   we are 
 led  to the following two  identities which are verified with the aid of Mathematica.
 
 {\bf Proposition 1.} {\em  
  For all $B_i$ and $z$
 \begin{equation}
   \label{3.11}
    H_{\sigma(s)}(z, (B_i) ) = H_s(z, (\hat{B}_i)),
  \end{equation}
  $s= 1, \dots, 6$.  }

 {\bf Proposition 2.} {\em  
   For all $B_i \neq 0$ and $z \neq 0$
  \begin{equation}
    \label{3.12}
     H_{s}(z, (B_i) ) = H_s^{as}(z, (B_i)) H_s(z^{-1}, (\hat{B}_i^{-1})),
   \end{equation}
   $s = 1, \dots, 6$. }
  
 We call  (\ref{3.11}) symmetry relations, and (\ref{3.12}) duality ones.
 
   \subsection{Exact solution for $E_6$, fluxes and asymptotics }
  
  The solution  (\ref{2.30})- (\ref{2.32a})  in our case reads
  
     \bear{3.13}
      g= \Bigl(\prod_{s = 1}^{6} H_s^{2 h /(D-2)} \Bigr)
      \biggl\{ d\rho \otimes d \rho  +
      \Bigl(\prod_{s = 1}^{6} H_s^{-2 h} \Bigr) \rho^2 d\phi \otimes d\phi +
        g^2  \biggr\},
     \\  \label{3.14}
      \exp(\varphi^a)=
      \prod_{s = 1}^{6} H_s^{h \lambda_{s}^a},
     \\  \label{3.15}
      F^s= {\cal B}^s \rho d\rho \wedge d \phi,
      \ear
     $a, s = 1,\dots, 6$, where  $g^1 = d \phi \otimes d \phi$ is a
      metric on $M_1 = S^1$ ($0 < \phi < 2 \pi$), 
       $g^2$ is a  Ricci-flat metric on $M_{2}$ of signature $(-,+,\dots,+)$.
   Here 
   \begin{equation}
       \label{3.16}
   {\cal B}^s =  - Q_s 
   \left( \prod_{l = 1}^{6}  H_{l}^{- A_{s l}} \right)
     \end{equation} 
  and due to (\ref{2.16})-(\ref{2.18})
    \begin{equation}
           \label{3.17}
    K = K_s =   \frac{D - 3}{D -2} +  \vec{\lambda}_{s}^2,
      \end{equation} 
    $h_s = h = K^{-1}$,    
      \beq{3.18}
        \vec{\lambda}_{s} \vec{\lambda}_{s'} = 
            \frac{1}{2} K A_{s s'}  - \frac{D - 3}{D -2} \equiv \Gamma_{s s'},
      \eeq    
     $s,s' = 1, \dots, 6$. For large enough $K$ there exist vectors 
     $\vec{\lambda}_{s}$ of equal length which obey relations  (\ref{3.18}).
     Indeed, the matrix $(\Gamma_{ss'})$ is positive-definite
     for $K > K_0$, where $K_0$ is some positive number. Hence there exists
     a matrix $\Lambda$, such that $\Lambda^{T}\Lambda = \Gamma$. We put
      $(\Lambda_{as}) = (\lambda_{s}^a)$ and get the set of vectors obeying
      (\ref{3.18}).
           
           {\bf Remark.} {\em   Let us  put $h_{\alpha\beta} = - \delta_{\alpha \beta}$.
          It may be shown (along a line as  was done for $h_{\alpha\beta} =\delta_{\alpha \beta}$)
           that, for $K < K_0 $,  where $K_0$ is some negative number,     
           there exist vectors  $\vec{\lambda}_{s}$ of equal length which obey relations  
          \beq{3.18r}
                  - \vec{\lambda}_{s} \vec{\lambda}_{s'} = 
                      \frac{1}{2} K A_{s s'}  - \frac{D - 3}{D -2},
                \eeq 
           following from (\ref{2.17}) and (\ref{2.18}).
           Thus, for both choices of signatures $h_{\alpha\beta} = \pm \delta_{\alpha \beta}$
           we get the same algebra (in our case $E_6$) and the same hidden group $G_T$. So,           
           the properties of the matrix $(h_{\alpha\beta})$ are not a priori known from the properties of the group
           $G_T$. 
           In the case of phantom scalar fields,  when $h_{\alpha\beta} = - \delta_{\alpha \beta}$,  we get 
            solutions which are defined for $\rho < \rho_0$, where $\rho_0 > 0$.
           The cosmological analogs of such solutions with phantom scalar fields 
          where considered for Lie algebras of rank $2$ and $3$ in refs. \cite{IKM-08} and   \cite{Gol-10},
          respectively. We note that another (sigma model) hidden group $G_{sm}$ (see Introduction)
           depends upon the choice of  the matrix $(h_{\alpha\beta})$ \cite{I-98}.}

      Now let us consider oriented $2$-dimensional manifold 
      $M_{*} =(0, + \infty)  \times S^1$. The flux integrals
       \beq{3.19}
       \Phi^s = \int_{M_{*}} F^s =
         2 \pi \int_{0}^{+ \infty} d \rho \rho {\cal B}^s ,
       \eeq    
      are convergent since due to 
      \beq{3.20}
       H_s \sim C_s \rho^{2n_s}, 
       \qquad C_s = \prod_{l = 1}^{6} B_l^{\nu^{sl}},
            \eeq 
      for $\rho \to + \infty$, and the equality $\sum_{1}^{6} A_{s l} n_l = 2$
      (following from (\ref{1.4})), we get
      \beq{3.21}
           {\cal B}^s \sim - Q_s C^s \rho^{-4},               
       \eeq 
      as $\rho \to + \infty$, where 
       \beq{3.22}
        C^s =  \prod_{l = 1}^{6} C_l^{-A_{sl}} = 
        \prod_{k = 1}^{6} \prod_{l = 1}^{6} B_l^{- A_{sk} \nu^{ks}},               
       \eeq 
      $s =1, \dots, 6$.   Due to (\ref{3.8})  we get $ A \nu = I + P$ 
        \beq{3.23}
              C^s =  \prod_{l = 1}^{6} B_l^{-(I+P)_{sl}} = 
               \prod_{l = 1}^{6} B_l^{- \delta^l_s - \delta^l_{\sigma(s)} } = 
               B_s^{-1} B_{\sigma(s)}^{-1},               
             \eeq 
      $s =1, \dots, 6$.
       
    By using the equations (\ref{1.1}) we obtain
      \bear{3.24}
      \int_{0}^{+ \infty} d \rho \rho {\cal B}^s = - Q_s P_s^{-1}
      \frac12 \int_{0}^{+ \infty} d z  
      \frac{d}{dz} \left( \frac{ z}{H_s} \frac{d}{dz} H_s \right)
      \\ \nonumber
      =   -  \frac12 Q_s P_s^{-1} \lim_{z \to + \infty}   
              \left( \frac{ z}{H_s} \frac{d}{dz} H_s \right) = -  \frac12 n_s Q_s P_s^{-1},          
      \ear
   which implies (see (\ref{2.21}))
            \beq{3.25}
         \Phi^s = -  4 \pi n_s Q_s^{-1} h,  \qquad  h = K^{-1}, 
         \eeq
   $s =1, \dots, 6$.
  
  It is remarkable that any  flux $\Phi^s$ depends  only upon $n_s$ 
  and the integration constant  $Q_s$, which for $D =4$ and 
  $g^2 = -  dt \otimes dt + d x \otimes d x$ is coinciding up to a sign with 
  the value of the $x$-component of the magnetic field on the axis of symmetry. 
  
  Analogous relations were found recently in ref. \cite{BolIvas-R2-17} for solutions 
  corresponding to  Lie algebras of rank $2$; see also ref. \cite{Ivas-Symmetry-17}.
      
  The asymptotic relations for the solution under consideration for 
  $\rho \to + \infty $ read
  
  \bear{3.26}
        g_{as} = \Bigl(\prod_{l = 1}^{6} B_l^{n_l} \Bigr)^{2 h /(D-2)} \rho^{2A}
        \biggl\{ d\rho \otimes d \rho  \qquad \\ \nonumber
         +
        \Bigl(\prod_{s = 1}^{6} B_s^{n_s} \Bigr)^{-2 h} 
        \rho^{2 - 2A (D-2)} d\phi \otimes d\phi +  g^2 \biggr\},
        \\  \label{3.27}
        \varphi^a_{as} = h \sum_{s = 1}^{6} \lambda_{s}^a 
        (\sum_{l = 1}^{6} \nu^{sl} \ln B_l + 2 n_s \ln \rho ),
        \\  \label{3.28}
        F^s_{as} = - Q_s  B_s^{-1} B_{\sigma(s)}^{-1} \rho^{-3}  d\rho \wedge d \phi,
        \ear       
        $a, s =1, \dots, 6$,
        where 
         \beq{3.25}
            A =  (2 h /(D-2)) \sum_{s = 1}^{6} n_{s} = (312 h /(D-2)). 
         \eeq
     In derivation of asymptotic relations eqs. (\ref{3.10a})  (\ref{3.20}), (\ref{3.21}) and (\ref{3.23}) 
     were used.

  \section{\bf Conclusions}

  Here we have obtained a multidimensional generalization of  Melvin's
  solution for the Lie algebra $E_6$.   The solution is governed by a set of six  fluxbrane polynomials $H_s(z)$, $s =1,\dots,6$, which are  presented in the appendix.  These polynomials  define special solutions to open Toda chain equations corresponding to the Lie algebra $E_6$.
   
   The polynomials $H_s(z)$ depend also upon parameters $Q_s$, which   
   are coinciding for $D =4$  (up to a sign) with the values  of  colored 
   magnetic fields on the axis of symmetry. The symmetry and duality identities for polynomials were verified.  
   The duality identities may be used  
   in deriving  $(1/\rho)$-expansion for solutions at large distances $\rho$, e.g. 
   for asymptotic relations,  which are presented   in the paper.  
   The power-law asymptotic relations for  $E_6$-polynomials at large $z$ are governed by   integer-valued matrix $\nu$. This matrix is related to the inverse Cartan matrix $A^{-1}$ by the formula  $\nu = A^{-1} (I + P)$, 
   where $I$ is identity matrix and  $P$ is permutation matrix. The matrix $P$ corresponds to a permutation 
   $\sigma \in S_6 $,   which is the generator of the $Z_2$-group  of symmetry of the Dynkin diagram.
  
  We have also calculated $2d$ flux integrals $\Phi^s$, $s =1, \dots, 6$.
  Any flux  $\Phi^s$ depends only upon one parameter $Q_s$, while the integrand 
  $F^s$ depends upon all parameters $Q_1, \dots, Q_6$. An open question is how to
  apply the approach of this paper to other finite-dimensional simple Lie algebras.

  \begin{center}
  {\bf Acknowledgments}
  \end{center}

 This work was supported in part by the Russian Foundation for
 Basic Research Grant no. 16-02-00602 and by the Ministry of Education of the Russian Federation (the
 agreement number 02.a03.21.0008 of 24 June 2016).
 
   \begin{center}
  {\bf Appendix}
   \end{center}

   In this appendix we present polynomials corresponding to the Lie algebra $E_6$.
   The polynomials were calculated by using a certain program in Mathematica.
   We denote the variable $z$ in bold and capital inside the polynomials for better readability:

\vspace{1em}

\newcounter{mathematicapage}
\tolerance7000
\small
\noindent\(\fbox{$H_1$}=B_1^2 B_2^3 B_3^4 B_4^3 B_5^2 B_6^2 \textbf{Z}^{16}+16 B_1^2 B_2^3 B_3^4 B_4^3 B_5 B_6^2 \textbf{Z}^{15}+120 B_1^2 B_2^3 B_3^4 B_4^2 B_5 B_6^2
\textbf{Z}^{14}+560 B_1^2 B_2^3 B_3^3 B_4^2 B_5 B_6^2 \textbf{Z}^{13}+(1050 B_1^2 B_2^2 B_4^2 B_5 B_6^2 B_3^3+770 B_1^2 B_2^3 B_4^2 B_5 B_6 B_3^3) \textbf{Z}^{12}+(672
B_1 B_2^2 B_4^2 B_5 B_6^2 B_3^3+3696 B_1^2 B_2^2 B_4^2 B_5 B_6 B_3^3) \textbf{Z}^{11}+(3696 B_1 B_2^2 B_4^2 B_5 B_6 B_3^3+4312 B_1^2 B_2^2 B_4^2
B_5 B_6 B_3^2) \textbf{Z}^{10}+(8800 B_1 B_2^2 B_4^2 B_5 B_6 B_3^2+2640 B_1^2 B_2^2 B_4 B_5 B_6 B_3^2) \textbf{Z}^9+(660 B_1^2 B_2^2 B_4 B_6
B_3^2+4125 B_1 B_2 B_4^2 B_5 B_6 B_3^2+8085 B_1 B_2^2 B_4 B_5 B_6 B_3^2) \textbf{Z}^8+(2640 B_1 B_2^2 B_4 B_6 B_3^2+8800 B_1 B_2 B_4 B_5 B_6 B_3^2)
\textbf{Z}^7+(4312 B_1 B_2 B_4 B_6 B_3^2+3696 B_1 B_2 B_4 B_5 B_6 B_3) \textbf{Z}^6+(672 B_1 B_2 B_3 B_4 B_5+3696 B_1 B_2 B_3 B_4 B_6) \textbf{Z}^5+(1050
B_1 B_2 B_3 B_4+770 B_1 B_2 B_3 B_6) \textbf{Z}^4+560 B_1 B_2 B_3 \textbf{Z}^3+120 B_1 B_2 \textbf{Z}^2+16 B_1 \textbf{Z}+1\)

\vspace{1em}
\noindent\(\fbox{$H_2$}=B_1^3 B_2^6 B_3^8 B_4^6 B_5^3 B_6^4 \textbf{Z}^{30}+30 B_1^3 B_2^6 B_3^8 B_4^5 B_5^3 B_6^4 \textbf{Z}^{29}+(120 B_1^3 B_2^6 B_4^5 B_5^2
B_6^4 B_3^8+315 B_1^3 B_2^6 B_4^5 B_5^3 B_6^4 B_3^7) \textbf{Z}^{28}+(1050 B_1^3 B_2^5 B_4^5 B_5^3 B_6^4 B_3^7+2240 B_1^3 B_2^6 B_4^5 B_5^2 B_6^4
B_3^7+770 B_1^3 B_2^6 B_4^5 B_5^3 B_6^3 B_3^7) \textbf{Z}^{27}+(1050 B_1^2 B_2^5 B_4^5 B_5^3 B_6^4 B_3^7+9450 B_1^3 B_2^5 B_4^5 B_5^2 B_6^4 B_3^7+4200
B_1^3 B_2^6 B_4^4 B_5^2 B_6^4 B_3^7+5775 B_1^3 B_2^5 B_4^5 B_5^3 B_6^3 B_3^7+6930 B_1^3 B_2^6 B_4^5 B_5^2 B_6^3 B_3^7) \textbf{Z}^{26}+(10752 B_1^2
B_2^5 B_4^5 B_5^2 B_6^4 B_3^7+31500 B_1^3 B_2^5 B_4^4 B_5^2 B_6^4 B_3^7+8316 B_1^2 B_2^5 B_4^5 B_5^3 B_6^3 B_3^7+59136 B_1^3 B_2^5 B_4^5 B_5^2 B_6^3
B_3^7+23100 B_1^3 B_2^6 B_4^4 B_5^2 B_6^3 B_3^7+9702 B_1^3 B_2^5 B_4^5 B_5^3 B_6^3 B_3^6) \textbf{Z}^{25}+(45360 B_1^2 B_2^5 B_4^4 B_5^2 B_6^4
B_3^7+92400 B_1^2 B_2^5 B_4^5 B_5^2 B_6^3 B_3^7+249480 B_1^3 B_2^5 B_4^4 B_5^2 B_6^3 B_3^7+36750 B_1^3 B_2^5 B_4^4 B_5^2 B_6^4 B_3^6+26950 B_1^2
B_2^5 B_4^5 B_5^3 B_6^3 B_3^6+8085 B_1^3 B_2^5 B_4^4 B_5^3 B_6^3 B_3^6+107800 B_1^3 B_2^5 B_4^5 B_5^2 B_6^3 B_3^6+26950 B_1^3 B_2^6 B_4^4 B_5^2 B_6^3
B_3^6) \textbf{Z}^{24}+(443520 B_1^2 B_2^5 B_4^4 B_5^2 B_6^3 B_3^7+94080 B_1^2 B_2^5 B_4^4 B_5^2 B_6^4 B_3^6+16500 B_1^2 B_2^4 B_4^5 B_5^3 B_6^3
B_3^6+32340 B_1^2 B_2^5 B_4^4 B_5^3 B_6^3 B_3^6+316800 B_1^2 B_2^5 B_4^5 B_5^2 B_6^3 B_3^6+1132560 B_1^3 B_2^5 B_4^4 B_5^2 B_6^3 B_3^6) \textbf{Z}^{23}+(44100
B_1^2 B_2^4 B_4^4 B_5^2 B_6^4 B_3^6+44550 B_1^2 B_2^4 B_4^4 B_5^3 B_6^3 B_3^6+202125 B_1^2 B_2^4 B_4^5 B_5^2 B_6^3 B_3^6+3256110 B_1^2 B_2^5 B_4^4
B_5^2 B_6^3 B_3^6+242550 B_1^3 B_2^4 B_4^4 B_5^2 B_6^3 B_3^6+495000 B_1^3 B_2^5 B_4^3 B_5^2 B_6^3 B_3^6+32340 B_1^3 B_2^5 B_4^4 B_5 B_6^3 B_3^6+177870
B_1^3 B_2^5 B_4^4 B_5^2 B_6^2 B_3^6+1358280 B_1^3 B_2^5 B_4^4 B_5^2 B_6^3 B_3^5) \textbf{Z}^{22}+(2674100 B_1^2 B_2^4 B_4^4 B_5^2 B_6^3 B_3^6+2182950
B_1^2 B_2^5 B_4^3 B_5^2 B_6^3 B_3^6+168960 B_1^2 B_2^5 B_4^4 B_5 B_6^3 B_3^6+178200 B_1^3 B_2^5 B_4^3 B_5 B_6^3 B_3^6+711480 B_1^2 B_2^5 B_4^4 B_5^2
B_6^2 B_3^6+23100 B_1^2 B_2^4 B_4^4 B_5^3 B_6^3 B_3^5+4928000 B_1^2 B_2^5 B_4^4 B_5^2 B_6^3 B_3^5+1131900 B_1^3 B_2^4 B_4^4 B_5^2 B_6^3 B_3^5+1478400
B_1^3 B_2^5 B_4^3 B_5^2 B_6^3 B_3^5+830060 B_1^3 B_2^5 B_4^4 B_5^2 B_6^2 B_3^5) \textbf{Z}^{21}+(155232 B_1 B_2^4 B_4^4 B_5^2 B_6^3 B_3^6+3234000
B_1^2 B_2^4 B_4^3 B_5^2 B_6^3 B_3^6+349272 B_1^2 B_2^4 B_4^4 B_5 B_6^3 B_3^6+970200 B_1^2 B_2^5 B_4^3 B_5 B_6^3 B_3^6+853776 B_1^2 B_2^4 B_4^4 B_5^2
B_6^2 B_3^6+9315306 B_1^2 B_2^4 B_4^4 B_5^2 B_6^3 B_3^5+7074375 B_1^2 B_2^5 B_4^3 B_5^2 B_6^3 B_3^5+1559250 B_1^3 B_2^4 B_4^3 B_5^2 B_6^3 B_3^5+577500
B_1^3 B_2^5 B_4^3 B_5 B_6^3 B_3^5+5082 B_1^2 B_2^4 B_4^4 B_5^3 B_6^2 B_3^5+3811500 B_1^2 B_2^5 B_4^4 B_5^2 B_6^2 B_3^5+996072 B_1^3 B_2^4 B_4^4 B_5^2
B_6^2 B_3^5+1143450 B_1^3 B_2^5 B_4^3 B_5^2 B_6^2 B_3^5) \textbf{Z}^{20}+(2069760 B_1^2 B_2^4 B_4^3 B_5 B_6^3 B_3^6+1478400 B_1 B_2^4 B_4^4 B_5^2
B_6^3 B_3^5+4331250 B_1^2 B_2^3 B_4^4 B_5^2 B_6^3 B_3^5+21801780 B_1^2 B_2^4 B_4^3 B_5^2 B_6^3 B_3^5+369600 B_1^2 B_2^4 B_4^4 B_5 B_6^3 B_3^5+3326400
B_1^2 B_2^5 B_4^3 B_5 B_6^3 B_3^5+693000 B_1^3 B_2^4 B_4^3 B_5 B_6^3 B_3^5+11384100 B_1^2 B_2^4 B_4^4 B_5^2 B_6^2 B_3^5+6225450 B_1^2 B_2^5 B_4^3
B_5^2 B_6^2 B_3^5+2439360 B_1^3 B_2^4 B_4^3 B_5^2 B_6^2 B_3^5+508200 B_1^3 B_2^5 B_4^3 B_5 B_6^2 B_3^5) \textbf{Z}^{19}+(1559250 B_1 B_2^3 B_4^4
B_5^2 B_6^3 B_3^5+3056130 B_1 B_2^4 B_4^3 B_5^2 B_6^3 B_3^5+14437500 B_1^2 B_2^3 B_4^3 B_5^2 B_6^3 B_3^5+14314300 B_1^2 B_2^4 B_4^3 B_5 B_6^3 B_3^5+2032800
B_1 B_2^4 B_4^4 B_5^2 B_6^2 B_3^5+8575875 B_1^2 B_2^3 B_4^4 B_5^2 B_6^2 B_3^5+28420210 B_1^2 B_2^4 B_4^3 B_5^2 B_6^2 B_3^5+127050 B_1^2 B_2^4 B_4^4
B_5 B_6^2 B_3^5+3176250 B_1^2 B_2^5 B_4^3 B_5 B_6^2 B_3^5+1372140 B_1^3 B_2^4 B_4^3 B_5 B_6^2 B_3^5+6338640 B_1^2 B_2^4 B_4^3 B_5^2 B_6^3 B_3^4+2371600
B_1^2 B_2^4 B_4^4 B_5^2 B_6^2 B_3^4+711480 B_1^3 B_2^4 B_4^3 B_5^2 B_6^2 B_3^4) \textbf{Z}^{18}+(5913600 B_1 B_2^3 B_4^3 B_5^2 B_6^3 B_3^5+1774080
B_1 B_2^4 B_4^3 B_5 B_6^3 B_3^5+10187100 B_1^2 B_2^3 B_4^3 B_5 B_6^3 B_3^5+577500 B_1^2 B_2^4 B_4^2 B_5 B_6^3 B_3^5+3811500 B_1 B_2^3 B_4^4 B_5^2
B_6^2 B_3^5+7470540 B_1 B_2^4 B_4^3 B_5^2 B_6^2 B_3^5+32524800 B_1^2 B_2^3 B_4^3 B_5^2 B_6^2 B_3^5+18705960 B_1^2 B_2^4 B_4^3 B_5 B_6^2 B_3^5+8731800
B_1^2 B_2^3 B_4^3 B_5^2 B_6^3 B_3^4+8279040 B_1^2 B_2^4 B_4^3 B_5 B_6^3 B_3^4+4446750 B_1^2 B_2^3 B_4^4 B_5^2 B_6^2 B_3^4+16625700 B_1^2 B_2^4 B_4^3
B_5^2 B_6^2 B_3^4+711480 B_1^3 B_2^4 B_4^3 B_5 B_6^2 B_3^4) \textbf{Z}^{17}+(4527600 B_1 B_2^3 B_4^3 B_5 B_6^3 B_3^5+18295200 B_1 B_2^3 B_4^3 B_5^2
B_6^2 B_3^5+5488560 B_1 B_2^4 B_4^3 B_5 B_6^2 B_3^5+24901800 B_1^2 B_2^3 B_4^3 B_5 B_6^2 B_3^5+508200 B_1^2 B_2^4 B_4^2 B_5 B_6^2 B_3^5+3880800 B_1
B_2^3 B_4^3 B_5^2 B_6^3 B_3^4+11884950 B_1^2 B_2^3 B_4^3 B_5 B_6^3 B_3^4+2182950 B_1^2 B_2^4 B_4^2 B_5 B_6^3 B_3^4+2268750 B_1 B_2^3 B_4^4 B_5^2
B_6^2 B_3^4+4446750 B_1 B_2^4 B_4^3 B_5^2 B_6^2 B_3^4+45530550 B_1^2 B_2^3 B_4^3 B_5^2 B_6^2 B_3^4+1334025 B_1^2 B_2^4 B_4^2 B_5^2 B_6^2 B_3^4+18478980
B_1^2 B_2^4 B_4^3 B_5 B_6^2 B_3^4+108900 B_1^3 B_2^4 B_4^2 B_5 B_6^2 B_3^4+1584660 B_1^2 B_2^4 B_4^3 B_5^2 B_6 B_3^4) \textbf{Z}^{16}+(15937152
B_1 B_2^3 B_4^3 B_5 B_6^2 B_3^5+5588352 B_1 B_2^3 B_4^3 B_5 B_6^3 B_3^4+3234000 B_1^2 B_2^3 B_4^2 B_5 B_6^3 B_3^4+34036496 B_1 B_2^3 B_4^3 B_5^2
B_6^2 B_3^4+5808000 B_1^2 B_2^3 B_4^2 B_5^2 B_6^2 B_3^4+5808000 B_1 B_2^4 B_4^3 B_5 B_6^2 B_3^4+53742416 B_1^2 B_2^3 B_4^3 B_5 B_6^2 B_3^4+6203600
B_1^2 B_2^4 B_4^2 B_5 B_6^2 B_3^4+5588352 B_1^2 B_2^3 B_4^3 B_5^2 B_6 B_3^4+3234000 B_1^2 B_2^4 B_4^3 B_5 B_6 B_3^4+15937152 B_1^2 B_2^3 B_4^3 B_5^2
B_6^2 B_3^3) \textbf{Z}^{15}+(108900 B_1^2 B_3^4 B_4^2 B_6^2 B_2^4+1334025 B_1 B_3^4 B_4^2 B_5 B_6^2 B_2^4+508200 B_1^2 B_3^3 B_4^2 B_5 B_6^2 B_2^4+2182950
B_1^2 B_3^4 B_4^2 B_5 B_6 B_2^4+1584660 B_1 B_3^4 B_4^2 B_5 B_6^3 B_2^3+18295200 B_1 B_3^3 B_4^3 B_5^2 B_6^2 B_2^3+4446750 B_1 B_3^4 B_4^2 B_5^2
B_6^2 B_2^3+5488560 B_1^2 B_3^3 B_4^2 B_5^2 B_6^2 B_2^3+45530550 B_1 B_3^4 B_4^3 B_5 B_6^2 B_2^3+24901800 B_1^2 B_3^3 B_4^3 B_5 B_6^2 B_2^3+18478980
B_1^2 B_3^4 B_4^2 B_5 B_6^2 B_2^3+3880800 B_1 B_3^4 B_4^3 B_5^2 B_6 B_2^3+4527600 B_1^2 B_3^3 B_4^3 B_5^2 B_6 B_2^3+11884950 B_1^2 B_3^4 B_4^3 B_5
B_6 B_2^3+2268750 B_1 B_3^4 B_4^3 B_5^2 B_6^2 B_2^2) \textbf{Z}^{14}+(577500 B_1^2 B_3^3 B_4^2 B_5 B_6 B_2^4+711480 B_1^2 B_3^4 B_4^2 B_6^2 B_2^3+7470540
B_1 B_3^3 B_4^2 B_5^2 B_6^2 B_2^3+32524800 B_1 B_3^3 B_4^3 B_5 B_6^2 B_2^3+16625700 B_1 B_3^4 B_4^2 B_5 B_6^2 B_2^3+18705960 B_1^2 B_3^3 B_4^2 B_5
B_6^2 B_2^3+5913600 B_1 B_3^3 B_4^3 B_5^2 B_6 B_2^3+1774080 B_1^2 B_3^3 B_4^2 B_5^2 B_6 B_2^3+8731800 B_1 B_3^4 B_4^3 B_5 B_6 B_2^3+10187100 B_1^2
B_3^3 B_4^3 B_5 B_6 B_2^3+8279040 B_1^2 B_3^4 B_4^2 B_5 B_6 B_2^3+3811500 B_1 B_3^3 B_4^3 B_5^2 B_6^2 B_2^2+4446750 B_1 B_3^4 B_4^3 B_5 B_6^2 B_2^2)
\textbf{Z}^{13}+(711480 B_1 B_2^3 B_4^2 B_6^2 B_3^4+2371600 B_1 B_2^2 B_4^2 B_5 B_6^2 B_3^4+6338640 B_1 B_2^3 B_4^2 B_5 B_6 B_3^4+1372140 B_1^2 B_2^3
B_4^2 B_6^2 B_3^3+2032800 B_1 B_2^2 B_4^2 B_5^2 B_6^2 B_3^3+8575875 B_1 B_2^2 B_4^3 B_5 B_6^2 B_3^3+28420210 B_1 B_2^3 B_4^2 B_5 B_6^2 B_3^3+127050
B_1^2 B_2^2 B_4^2 B_5 B_6^2 B_3^3+3176250 B_1^2 B_2^3 B_4 B_5 B_6^2 B_3^3+1559250 B_1 B_2^2 B_4^3 B_5^2 B_6 B_3^3+3056130 B_1 B_2^3 B_4^2 B_5^2 B_6
B_3^3+14437500 B_1 B_2^3 B_4^3 B_5 B_6 B_3^3+14314300 B_1^2 B_2^3 B_4^2 B_5 B_6 B_3^3) \textbf{Z}^{12}+(2439360 B_1 B_3^3 B_4^2 B_6^2 B_2^3+508200
B_1^2 B_3^3 B_4 B_6^2 B_2^3+6225450 B_1 B_3^3 B_4 B_5 B_6^2 B_2^3+693000 B_1^2 B_3^3 B_4^2 B_6 B_2^3+21801780 B_1 B_3^3 B_4^2 B_5 B_6 B_2^3+2069760
B_1^2 B_3^2 B_4^2 B_5 B_6 B_2^3+3326400 B_1^2 B_3^3 B_4 B_5 B_6 B_2^3+11384100 B_1 B_3^3 B_4^2 B_5 B_6^2 B_2^2+1478400 B_1 B_3^3 B_4^2 B_5^2 B_6
B_2^2+4331250 B_1 B_3^3 B_4^3 B_5 B_6 B_2^2+369600 B_1^2 B_3^3 B_4^2 B_5 B_6 B_2^2) \textbf{Z}^{11}+(1143450 B_1 B_3^3 B_4 B_6^2 B_2^3+1559250
B_1 B_3^3 B_4^2 B_6 B_2^3+577500 B_1^2 B_3^3 B_4 B_6 B_2^3+3234000 B_1 B_3^2 B_4^2 B_5 B_6 B_2^3+7074375 B_1 B_3^3 B_4 B_5 B_6 B_2^3+970200 B_1^2
B_3^2 B_4 B_5 B_6 B_2^3+996072 B_1 B_3^3 B_4^2 B_6^2 B_2^2+5082 B_3^3 B_4^2 B_5 B_6^2 B_2^2+853776 B_1 B_3^2 B_4^2 B_5 B_6^2 B_2^2+3811500 B_1 B_3^3
B_4 B_5 B_6^2 B_2^2+155232 B_1 B_3^2 B_4^2 B_5^2 B_6 B_2^2+9315306 B_1 B_3^3 B_4^2 B_5 B_6 B_2^2+349272 B_1^2 B_3^2 B_4^2 B_5 B_6 B_2^2) \textbf{Z}^{10}+(1478400
B_1 B_3^3 B_4 B_6 B_2^3+178200 B_1^2 B_3^2 B_4 B_6 B_2^3+2182950 B_1 B_3^2 B_4 B_5 B_6 B_2^3+830060 B_1 B_3^3 B_4 B_6^2 B_2^2+711480 B_1 B_3^2 B_4
B_5 B_6^2 B_2^2+1131900 B_1 B_3^3 B_4^2 B_6 B_2^2+23100 B_3^3 B_4^2 B_5 B_6 B_2^2+2674100 B_1 B_3^2 B_4^2 B_5 B_6 B_2^2+4928000 B_1 B_3^3 B_4 B_5
B_6 B_2^2+168960 B_1^2 B_3^2 B_4 B_5 B_6 B_2^2) \textbf{Z}^9+(495000 B_1 B_3^2 B_4 B_6 B_2^3+177870 B_1 B_3^2 B_4 B_6^2 B_2^2+44100 B_1 B_3^2 B_4^2
B_5 B_2^2+242550 B_1 B_3^2 B_4^2 B_6 B_2^2+1358280 B_1 B_3^3 B_4 B_6 B_2^2+32340 B_1^2 B_3^2 B_4 B_6 B_2^2+44550 B_3^2 B_4^2 B_5 B_6 B_2^2+3256110
B_1 B_3^2 B_4 B_5 B_6 B_2^2+202125 B_1 B_3^2 B_4^2 B_5 B_6 B_2) \textbf{Z}^8+(94080 B_1 B_3^2 B_4 B_5 B_2^2+1132560 B_1 B_3^2 B_4 B_6 B_2^2+32340
B_3^2 B_4 B_5 B_6 B_2^2+443520 B_1 B_3 B_4 B_5 B_6 B_2^2+16500 B_3^2 B_4^2 B_5 B_6 B_2+316800 B_1 B_3^2 B_4 B_5 B_6 B_2) \textbf{Z}^7+(36750 B_1
B_3^2 B_4 B_2^2+45360 B_1 B_3 B_4 B_5 B_2^2+26950 B_1 B_3^2 B_6 B_2^2+8085 B_3^2 B_4 B_6 B_2^2+249480 B_1 B_3 B_4 B_6 B_2^2+107800 B_1 B_3^2 B_4
B_6 B_2+26950 B_3^2 B_4 B_5 B_6 B_2+92400 B_1 B_3 B_4 B_5 B_6 B_2) \textbf{Z}^6+(31500 B_1 B_3 B_4 B_2^2+23100 B_1 B_3 B_6 B_2^2+10752 B_1 B_3
B_4 B_5 B_2+9702 B_3^2 B_4 B_6 B_2+59136 B_1 B_3 B_4 B_6 B_2+8316 B_3 B_4 B_5 B_6 B_2) \textbf{Z}^5+(4200 B_1 B_3 B_2^2+9450 B_1 B_3 B_4 B_2+1050
B_3 B_4 B_5 B_2+6930 B_1 B_3 B_6 B_2+5775 B_3 B_4 B_6 B_2) \textbf{Z}^4+(2240 B_1 B_2 B_3+1050 B_2 B_4 B_3+770 B_2 B_6 B_3) \textbf{Z}^3+(120
B_1 B_2+315 B_3 B_2) \textbf{Z}^2+30 B_2 \textbf{Z}+1\)

\vspace{1em}
\noindent\(\fbox{$H_3$}=B_1^4 B_2^8 B_3^{12} B_4^8 B_5^4 B_6^6 \textbf{Z}^{42}+42 B_1^4 B_2^8 B_3^{11} B_4^8 B_5^4 B_6^6 \textbf{Z}^{41}+(315 B_1^4 B_2^7 B_4^8
B_5^4 B_6^6 B_3^{11}+315 B_1^4 B_2^8 B_4^7 B_5^4 B_6^6 B_3^{11}+231 B_1^4 B_2^8 B_4^8 B_5^4 B_6^5 B_3^{11}) \textbf{Z}^{40}+(560 B_1^3 B_2^7 B_4^8
B_5^4 B_6^6 B_3^{11}+4200 B_1^4 B_2^7 B_4^7 B_5^4 B_6^6 B_3^{11}+560 B_1^4 B_2^8 B_4^7 B_5^3 B_6^6 B_3^{11}+3080 B_1^4 B_2^7 B_4^8 B_5^4 B_6^5 B_3^{11}+3080
B_1^4 B_2^8 B_4^7 B_5^4 B_6^5 B_3^{11}) \textbf{Z}^{39}+(9450 B_1^3 B_2^7 B_4^7 B_5^4 B_6^6 B_3^{11}+9450 B_1^4 B_2^7 B_4^7 B_5^3 B_6^6 B_3^{11}+6930
B_1^3 B_2^7 B_4^8 B_5^4 B_6^5 B_3^{11}+51975 B_1^4 B_2^7 B_4^7 B_5^4 B_6^5 B_3^{11}+6930 B_1^4 B_2^8 B_4^7 B_5^3 B_6^5 B_3^{11}+11025 B_1^4 B_2^7
B_4^7 B_5^4 B_6^6 B_3^{10}+8085 B_1^4 B_2^7 B_4^8 B_5^4 B_6^5 B_3^{10}+8085 B_1^4 B_2^8 B_4^7 B_5^4 B_6^5 B_3^{10}) \textbf{Z}^{38}+(24192 B_1^3
B_2^7 B_4^7 B_5^3 B_6^6 B_3^{11}+133056 B_1^3 B_2^7 B_4^7 B_5^4 B_6^5 B_3^{11}+133056 B_1^4 B_2^7 B_4^7 B_5^3 B_6^5 B_3^{11}+44100 B_1^3 B_2^7 B_4^7
B_5^4 B_6^6 B_3^{10}+44100 B_1^4 B_2^7 B_4^7 B_5^3 B_6^6 B_3^{10}+32340 B_1^3 B_2^7 B_4^8 B_5^4 B_6^5 B_3^{10}+407484 B_1^4 B_2^7 B_4^7 B_5^4 B_6^5
B_3^{10}+32340 B_1^4 B_2^8 B_4^7 B_5^3 B_6^5 B_3^{10}) \textbf{Z}^{37}+(369600 B_1^3 B_2^7 B_4^7 B_5^3 B_6^5 B_3^{11}+36750 B_1^3 B_2^6 B_4^7 B_5^4
B_6^6 B_3^{10}+200704 B_1^3 B_2^7 B_4^7 B_5^3 B_6^6 B_3^{10}+36750 B_1^4 B_2^7 B_4^6 B_5^3 B_6^6 B_3^{10}+26950 B_1^3 B_2^6 B_4^8 B_5^4 B_6^5 B_3^{10}+1539384
B_1^3 B_2^7 B_4^7 B_5^4 B_6^5 B_3^{10}+202125 B_1^4 B_2^6 B_4^7 B_5^4 B_6^5 B_3^{10}+202125 B_1^4 B_2^7 B_4^6 B_5^4 B_6^5 B_3^{10}+1539384 B_1^4
B_2^7 B_4^7 B_5^3 B_6^5 B_3^{10}+26950 B_1^4 B_2^8 B_4^6 B_5^3 B_6^5 B_3^{10}+148225 B_1^4 B_2^7 B_4^7 B_5^4 B_6^4 B_3^{10}+916839 B_1^4 B_2^7 B_4^7
B_5^4 B_6^5 B_3^9) \textbf{Z}^{36}+(211680 B_1^3 B_2^6 B_4^7 B_5^3 B_6^6 B_3^{10}+211680 B_1^3 B_2^7 B_4^6 B_5^3 B_6^6 B_3^{10}+1853280 B_1^3 B_2^6
B_4^7 B_5^4 B_6^5 B_3^{10}+1164240 B_1^3 B_2^7 B_4^6 B_5^4 B_6^5 B_3^{10}+6044544 B_1^3 B_2^7 B_4^7 B_5^3 B_6^5 B_3^{10}+1164240 B_1^4 B_2^6 B_4^7
B_5^3 B_6^5 B_3^{10}+1853280 B_1^4 B_2^7 B_4^6 B_5^3 B_6^5 B_3^{10}+853776 B_1^3 B_2^7 B_4^7 B_5^4 B_6^4 B_3^{10}+853776 B_1^4 B_2^7 B_4^7 B_5^3
B_6^4 B_3^{10}+4527600 B_1^3 B_2^7 B_4^7 B_5^4 B_6^5 B_3^9+1358280 B_1^4 B_2^6 B_4^7 B_5^4 B_6^5 B_3^9+1358280 B_1^4 B_2^7 B_4^6 B_5^4 B_6^5 B_3^9+4527600
B_1^4 B_2^7 B_4^7 B_5^3 B_6^5 B_3^9+996072 B_1^4 B_2^7 B_4^7 B_5^4 B_6^4 B_3^9) \textbf{Z}^{35}+(396900 B_1^3 B_2^6 B_4^6 B_5^3 B_6^6 B_3^{10}+291060
B_1^2 B_2^6 B_4^7 B_5^4 B_6^5 B_3^{10}+2182950 B_1^3 B_2^6 B_4^6 B_5^4 B_6^5 B_3^{10}+9168390 B_1^3 B_2^6 B_4^7 B_5^3 B_6^5 B_3^{10}+9168390 B_1^3
B_2^7 B_4^6 B_5^3 B_6^5 B_3^{10}+2182950 B_1^4 B_2^6 B_4^6 B_5^3 B_6^5 B_3^{10}+291060 B_1^4 B_2^7 B_4^6 B_5^2 B_6^5 B_3^{10}+1600830 B_1^3 B_2^6
B_4^7 B_5^4 B_6^4 B_3^{10}+5336100 B_1^3 B_2^7 B_4^7 B_5^3 B_6^4 B_3^{10}+1600830 B_1^4 B_2^7 B_4^6 B_5^3 B_6^4 B_3^{10}+13222440 B_1^3 B_2^6 B_4^7
B_5^4 B_6^5 B_3^9+8489250 B_1^3 B_2^7 B_4^6 B_5^4 B_6^5 B_3^9+2546775 B_1^4 B_2^6 B_4^6 B_5^4 B_6^5 B_3^9+23654400 B_1^3 B_2^7 B_4^7 B_5^3 B_6^5
B_3^9+8489250 B_1^4 B_2^6 B_4^7 B_5^3 B_6^5 B_3^9+13222440 B_1^4 B_2^7 B_4^6 B_5^3 B_6^5 B_3^9+6225450 B_1^3 B_2^7 B_4^7 B_5^4 B_6^4 B_3^9+1867635
B_1^4 B_2^6 B_4^7 B_5^4 B_6^4 B_3^9+1867635 B_1^4 B_2^7 B_4^6 B_5^4 B_6^4 B_3^9+6225450 B_1^4 B_2^7 B_4^7 B_5^3 B_6^4 B_3^9) \textbf{Z}^{34}+(2069760
B_1^2 B_2^6 B_4^7 B_5^3 B_6^5 B_3^{10}+24147200 B_1^3 B_2^6 B_4^6 B_5^3 B_6^5 B_3^{10}+2069760 B_1^3 B_2^7 B_4^6 B_5^2 B_6^5 B_3^{10}+11383680 B_1^3
B_2^6 B_4^7 B_5^3 B_6^4 B_3^{10}+11383680 B_1^3 B_2^7 B_4^6 B_5^3 B_6^4 B_3^{10}+205800 B_1^3 B_2^6 B_4^6 B_5^3 B_6^6 B_3^9+3773000 B_1^2 B_2^6 B_4^7
B_5^4 B_6^5 B_3^9+9240000 B_1^3 B_2^5 B_4^7 B_5^4 B_6^5 B_3^9+37560600 B_1^3 B_2^6 B_4^6 B_5^4 B_6^5 B_3^9+82222140 B_1^3 B_2^6 B_4^7 B_5^3 B_6^5
B_3^9+82222140 B_1^3 B_2^7 B_4^6 B_5^3 B_6^5 B_3^9+37560600 B_1^4 B_2^6 B_4^6 B_5^3 B_6^5 B_3^9+9240000 B_1^4 B_2^7 B_4^5 B_5^3 B_6^5 B_3^9+3773000
B_1^4 B_2^7 B_4^6 B_5^2 B_6^5 B_3^9+27544440 B_1^3 B_2^6 B_4^7 B_5^4 B_6^4 B_3^9+13280960 B_1^3 B_2^7 B_4^6 B_5^4 B_6^4 B_3^9+6225450 B_1^4 B_2^6
B_4^6 B_5^4 B_6^4 B_3^9+41164200 B_1^3 B_2^7 B_4^7 B_5^3 B_6^4 B_3^9+13280960 B_1^4 B_2^6 B_4^7 B_5^3 B_6^4 B_3^9+27544440 B_1^4 B_2^7 B_4^6 B_5^3
B_6^4 B_3^9) \textbf{Z}^{33}+(5588352 B_1^2 B_2^6 B_4^6 B_5^3 B_6^5 B_3^{10}+5588352 B_1^3 B_2^6 B_4^6 B_5^2 B_6^5 B_3^{10}+30735936 B_1^3 B_2^6
B_4^6 B_5^3 B_6^4 B_3^{10}+5197500 B_1^2 B_2^5 B_4^7 B_5^4 B_6^5 B_3^9+10187100 B_1^2 B_2^6 B_4^6 B_5^4 B_6^5 B_3^9+38981250 B_1^3 B_2^5 B_4^6 B_5^4
B_6^5 B_3^9+28385280 B_1^2 B_2^6 B_4^7 B_5^3 B_6^5 B_3^9+63669375 B_1^3 B_2^5 B_4^7 B_5^3 B_6^5 B_3^9+440527626 B_1^3 B_2^6 B_4^6 B_5^3 B_6^5 B_3^9+63669375
B_1^3 B_2^7 B_4^5 B_5^3 B_6^5 B_3^9+38981250 B_1^4 B_2^6 B_4^5 B_5^3 B_6^5 B_3^9+28385280 B_1^3 B_2^7 B_4^6 B_5^2 B_6^5 B_3^9+10187100 B_1^4 B_2^6
B_4^6 B_5^2 B_6^5 B_3^9+5197500 B_1^4 B_2^7 B_4^5 B_5^2 B_6^5 B_3^9+7470540 B_1^2 B_2^6 B_4^7 B_5^4 B_6^4 B_3^9+28586250 B_1^3 B_2^5 B_4^7 B_5^4
B_6^4 B_3^9+87268104 B_1^3 B_2^6 B_4^6 B_5^4 B_6^4 B_3^9+202848030 B_1^3 B_2^6 B_4^7 B_5^3 B_6^4 B_3^9+202848030 B_1^3 B_2^7 B_4^6 B_5^3 B_6^4 B_3^9+87268104
B_1^4 B_2^6 B_4^6 B_5^3 B_6^4 B_3^9+28586250 B_1^4 B_2^7 B_4^5 B_5^3 B_6^4 B_3^9+7470540 B_1^4 B_2^7 B_4^6 B_5^2 B_6^4 B_3^9+11884950 B_1^3 B_2^6
B_4^6 B_5^4 B_6^5 B_3^8+11884950 B_1^4 B_2^6 B_4^6 B_5^3 B_6^5 B_3^8+8715630 B_1^3 B_2^6 B_4^7 B_5^4 B_6^4 B_3^8+2614689 B_1^4 B_2^6 B_4^6 B_5^4
B_6^4 B_3^8+8715630 B_1^4 B_2^7 B_4^6 B_5^3 B_6^4 B_3^8) \textbf{Z}^{32}+(24948000 B_1^2 B_2^5 B_4^6 B_5^4 B_6^5 B_3^9+40748400 B_1^2 B_2^5 B_4^7
B_5^3 B_6^5 B_3^9+177031008 B_1^2 B_2^6 B_4^6 B_5^3 B_6^5 B_3^9+467082000 B_1^3 B_2^5 B_4^6 B_5^3 B_6^5 B_3^9+467082000 B_1^3 B_2^6 B_4^5 B_5^3 B_6^5
B_3^9+177031008 B_1^3 B_2^6 B_4^6 B_5^2 B_6^5 B_3^9+40748400 B_1^3 B_2^7 B_4^5 B_5^2 B_6^5 B_3^9+24948000 B_1^4 B_2^6 B_4^5 B_5^2 B_6^5 B_3^9+18295200
B_1^2 B_2^5 B_4^7 B_5^4 B_6^4 B_3^9+35858592 B_1^2 B_2^6 B_4^6 B_5^4 B_6^4 B_3^9+137214000 B_1^3 B_2^5 B_4^6 B_5^4 B_6^4 B_3^9+60984000 B_1^2 B_2^6
B_4^7 B_5^3 B_6^4 B_3^9+224116200 B_1^3 B_2^5 B_4^7 B_5^3 B_6^4 B_3^9+1339753968 B_1^3 B_2^6 B_4^6 B_5^3 B_6^4 B_3^9+224116200 B_1^3 B_2^7 B_4^5
B_5^3 B_6^4 B_3^9+137214000 B_1^4 B_2^6 B_4^5 B_5^3 B_6^4 B_3^9+60984000 B_1^3 B_2^7 B_4^6 B_5^2 B_6^4 B_3^9+35858592 B_1^4 B_2^6 B_4^6 B_5^2 B_6^4
B_3^9+18295200 B_1^4 B_2^7 B_4^5 B_5^2 B_6^4 B_3^9+29106000 B_1^3 B_2^5 B_4^6 B_5^4 B_6^5 B_3^8+191866752 B_1^3 B_2^6 B_4^6 B_5^3 B_6^5 B_3^8+29106000
B_1^4 B_2^6 B_4^5 B_5^3 B_6^5 B_3^8+21344400 B_1^3 B_2^5 B_4^7 B_5^4 B_6^4 B_3^8+66594528 B_1^3 B_2^6 B_4^6 B_5^4 B_6^4 B_3^8+71148000 B_1^3 B_2^6
B_4^7 B_5^3 B_6^4 B_3^8+71148000 B_1^3 B_2^7 B_4^6 B_5^3 B_6^4 B_3^8+66594528 B_1^4 B_2^6 B_4^6 B_5^3 B_6^4 B_3^8+21344400 B_1^4 B_2^7 B_4^5 B_5^3
B_6^4 B_3^8) \textbf{Z}^{31}+(353089660 B_1^2 B_2^5 B_4^6 B_5^3 B_6^5 B_3^9+247546530 B_1^2 B_2^6 B_4^5 B_5^3 B_6^5 B_3^9+707437500 B_1^3 B_2^5
B_4^5 B_5^3 B_6^5 B_3^9+60555264 B_1^2 B_2^6 B_4^6 B_5^2 B_6^5 B_3^9+247546530 B_1^3 B_2^5 B_4^6 B_5^2 B_6^5 B_3^9+353089660 B_1^3 B_2^6 B_4^5 B_5^2
B_6^5 B_3^9+111143340 B_1^2 B_2^5 B_4^6 B_5^4 B_6^4 B_3^9+155636250 B_1^2 B_2^5 B_4^7 B_5^3 B_6^4 B_3^9+542666124 B_1^2 B_2^6 B_4^6 B_5^3 B_6^4 B_3^9+1941993130
B_1^3 B_2^5 B_4^6 B_5^3 B_6^4 B_3^9+1941993130 B_1^3 B_2^6 B_4^5 B_5^3 B_6^4 B_3^9+542666124 B_1^3 B_2^6 B_4^6 B_5^2 B_6^4 B_3^9+155636250 B_1^3
B_2^7 B_4^5 B_5^2 B_6^4 B_3^9+111143340 B_1^4 B_2^6 B_4^5 B_5^2 B_6^4 B_3^9+5478396 B_1^3 B_2^6 B_4^6 B_5^3 B_6^3 B_3^9+20212500 B_1^2 B_2^5 B_4^6
B_5^4 B_6^5 B_3^8+110387200 B_1^2 B_2^6 B_4^6 B_5^3 B_6^5 B_3^8+388031490 B_1^3 B_2^5 B_4^6 B_5^3 B_6^5 B_3^8+388031490 B_1^3 B_2^6 B_4^5 B_5^3 B_6^5
B_3^8+110387200 B_1^3 B_2^6 B_4^6 B_5^2 B_6^5 B_3^8+20212500 B_1^4 B_2^6 B_4^5 B_5^2 B_6^5 B_3^8+14822500 B_1^2 B_2^5 B_4^7 B_5^4 B_6^4 B_3^8+29052100
B_1^2 B_2^6 B_4^6 B_5^4 B_6^4 B_3^8+253998360 B_1^3 B_2^5 B_4^6 B_5^4 B_6^4 B_3^8+8715630 B_1^3 B_2^6 B_4^5 B_5^4 B_6^4 B_3^8+181575625 B_1^3 B_2^5
B_4^7 B_5^3 B_6^4 B_3^8+1554121926 B_1^3 B_2^6 B_4^6 B_5^3 B_6^4 B_3^8+8715630 B_1^4 B_2^5 B_4^6 B_5^3 B_6^4 B_3^8+181575625 B_1^3 B_2^7 B_4^5 B_5^3
B_6^4 B_3^8+253998360 B_1^4 B_2^6 B_4^5 B_5^3 B_6^4 B_3^8+29052100 B_1^4 B_2^6 B_4^6 B_5^2 B_6^4 B_3^8+14822500 B_1^4 B_2^7 B_4^5 B_5^2 B_6^4 B_3^8+6391462
B_1^3 B_2^6 B_4^6 B_5^4 B_6^3 B_3^8+6391462 B_1^4 B_2^6 B_4^6 B_5^3 B_6^3 B_3^8) \textbf{Z}^{30}+(651974400 B_1^2 B_2^5 B_4^5 B_5^3 B_6^5 B_3^9+195592320
B_1^2 B_2^5 B_4^6 B_5^2 B_6^5 B_3^9+195592320 B_1^2 B_2^6 B_4^5 B_5^2 B_6^5 B_3^9+651974400 B_1^3 B_2^5 B_4^5 B_5^2 B_6^5 B_3^9+1644128640 B_1^2
B_2^5 B_4^6 B_5^3 B_6^4 B_3^9+1075757760 B_1^2 B_2^6 B_4^5 B_5^3 B_6^4 B_3^9+3585859200 B_1^3 B_2^5 B_4^5 B_5^3 B_6^4 B_3^9+292723200 B_1^2 B_2^6
B_4^6 B_5^2 B_6^4 B_3^9+1075757760 B_1^3 B_2^5 B_4^6 B_5^2 B_6^4 B_3^9+1644128640 B_1^3 B_2^6 B_4^5 B_5^2 B_6^4 B_3^9+413887320 B_1^2 B_2^5 B_4^6
B_5^3 B_6^5 B_3^8+356548500 B_1^2 B_2^6 B_4^5 B_5^3 B_6^5 B_3^8+1189465200 B_1^3 B_2^5 B_4^5 B_5^3 B_6^5 B_3^8+356548500 B_1^3 B_2^5 B_4^6 B_5^2
B_6^5 B_3^8+413887320 B_1^3 B_2^6 B_4^5 B_5^2 B_6^5 B_3^8+268939440 B_1^2 B_2^5 B_4^6 B_5^4 B_6^4 B_3^8+48024900 B_1^3 B_2^5 B_4^5 B_5^4 B_6^4 B_3^8+133402500
B_1^2 B_2^5 B_4^7 B_5^3 B_6^4 B_3^8+672348600 B_1^2 B_2^6 B_4^6 B_5^3 B_6^4 B_3^8+4320547560 B_1^3 B_2^5 B_4^6 B_5^3 B_6^4 B_3^8+4320547560 B_1^3
B_2^6 B_4^5 B_5^3 B_6^4 B_3^8+48024900 B_1^4 B_2^5 B_4^5 B_5^3 B_6^4 B_3^8+672348600 B_1^3 B_2^6 B_4^6 B_5^2 B_6^4 B_3^8+133402500 B_1^3 B_2^7 B_4^5
B_5^2 B_6^4 B_3^8+268939440 B_1^4 B_2^6 B_4^5 B_5^2 B_6^4 B_3^8+35218260 B_1^3 B_2^5 B_4^6 B_5^4 B_6^3 B_3^8+180457200 B_1^3 B_2^6 B_4^6 B_5^3 B_6^3
B_3^8+35218260 B_1^4 B_2^6 B_4^5 B_5^3 B_6^3 B_3^8+119528640 B_1^3 B_2^5 B_4^6 B_5^4 B_6^4 B_3^7+398428800 B_1^3 B_2^6 B_4^6 B_5^3 B_6^4 B_3^7+119528640
B_1^4 B_2^6 B_4^5 B_5^3 B_6^4 B_3^7) \textbf{Z}^{29}+(651974400 B_1^2 B_2^5 B_4^5 B_5^2 B_6^5 B_3^9+3585859200 B_1^2 B_2^5 B_4^5 B_5^3 B_6^4 B_3^9+1075757760
B_1^2 B_2^5 B_4^6 B_5^2 B_6^4 B_3^9+1075757760 B_1^2 B_2^6 B_4^5 B_5^2 B_6^4 B_3^9+3585859200 B_1^3 B_2^5 B_4^5 B_5^2 B_6^4 B_3^9+54573750 B_1^2
B_2^4 B_4^6 B_5^3 B_6^5 B_3^8+1671169500 B_1^2 B_2^5 B_4^5 B_5^3 B_6^5 B_3^8+298045440 B_1^2 B_2^5 B_4^6 B_5^2 B_6^5 B_3^8+298045440 B_1^2 B_2^6
B_4^5 B_5^2 B_6^5 B_3^8+1671169500 B_1^3 B_2^5 B_4^5 B_5^2 B_6^5 B_3^8+54573750 B_1^3 B_2^6 B_4^4 B_5^2 B_6^5 B_3^8+24502500 B_1^2 B_2^4 B_4^6 B_5^4
B_6^4 B_3^8+48024900 B_1^2 B_2^5 B_4^5 B_5^4 B_6^4 B_3^8+4609356210 B_1^2 B_2^5 B_4^6 B_5^3 B_6^4 B_3^8+300155625 B_1^3 B_2^4 B_4^6 B_5^3 B_6^4 B_3^8+3054383640
B_1^2 B_2^6 B_4^5 B_5^3 B_6^4 B_3^8+14283282150 B_1^3 B_2^5 B_4^5 B_5^3 B_6^4 B_3^8+300155625 B_1^3 B_2^6 B_4^4 B_5^3 B_6^4 B_3^8+446054400 B_1^2
B_2^6 B_4^6 B_5^2 B_6^4 B_3^8+3054383640 B_1^3 B_2^5 B_4^6 B_5^2 B_6^4 B_3^8+4609356210 B_1^3 B_2^6 B_4^5 B_5^2 B_6^4 B_3^8+48024900 B_1^4 B_2^5
B_4^5 B_5^2 B_6^4 B_3^8+24502500 B_1^4 B_2^6 B_4^4 B_5^2 B_6^4 B_3^8+35218260 B_1^2 B_2^5 B_4^6 B_5^4 B_6^3 B_3^8+117394200 B_1^2 B_2^6 B_4^6 B_5^3
B_6^3 B_3^8+607296690 B_1^3 B_2^5 B_4^6 B_5^3 B_6^3 B_3^8+607296690 B_1^3 B_2^6 B_4^5 B_5^3 B_6^3 B_3^8+117394200 B_1^3 B_2^6 B_4^6 B_5^2 B_6^3 B_3^8+35218260
B_1^4 B_2^6 B_4^5 B_5^2 B_6^3 B_3^8+499167900 B_1^3 B_2^5 B_4^5 B_5^3 B_6^5 B_3^7+186763500 B_1^2 B_2^5 B_4^6 B_5^4 B_6^4 B_3^7+56029050 B_1^3 B_2^5
B_4^5 B_5^4 B_6^4 B_3^7+2655776970 B_1^3 B_2^5 B_4^6 B_5^3 B_6^4 B_3^7+2655776970 B_1^3 B_2^6 B_4^5 B_5^3 B_6^4 B_3^7+56029050 B_1^4 B_2^5 B_4^5
B_5^3 B_6^4 B_3^7+186763500 B_1^4 B_2^6 B_4^5 B_5^2 B_6^4 B_3^7+41087970 B_1^3 B_2^5 B_4^6 B_5^4 B_6^3 B_3^7+136959900 B_1^3 B_2^6 B_4^6 B_5^3 B_6^3
B_3^7+41087970 B_1^4 B_2^6 B_4^5 B_5^3 B_6^3 B_3^7) \textbf{Z}^{28}+(4079910912 B_1^2 B_2^5 B_4^5 B_5^2 B_6^4 B_3^9+323400000 B_1^2 B_2^4 B_4^5
B_5^3 B_6^5 B_3^8+2253071744 B_1^2 B_2^5 B_4^5 B_5^2 B_6^5 B_3^8+323400000 B_1^3 B_2^5 B_4^4 B_5^2 B_6^5 B_3^8+11383680 B_1 B_2^5 B_4^6 B_5^3 B_6^4
B_3^8+830060000 B_1^2 B_2^4 B_4^6 B_5^3 B_6^4 B_3^8+18476731056 B_1^2 B_2^5 B_4^5 B_5^3 B_6^4 B_3^8+1778700000 B_1^3 B_2^4 B_4^5 B_5^3 B_6^4 B_3^8+1778700000
B_1^3 B_2^5 B_4^4 B_5^3 B_6^4 B_3^8+4063973760 B_1^2 B_2^5 B_4^6 B_5^2 B_6^4 B_3^8+4063973760 B_1^2 B_2^6 B_4^5 B_5^2 B_6^4 B_3^8+18476731056 B_1^3
B_2^5 B_4^5 B_5^2 B_6^4 B_3^8+830060000 B_1^3 B_2^6 B_4^4 B_5^2 B_6^4 B_3^8+11383680 B_1^3 B_2^6 B_4^5 B_5 B_6^4 B_3^8+871627680 B_1^2 B_2^5 B_4^6
B_5^3 B_6^3 B_3^8+766975440 B_1^2 B_2^6 B_4^5 B_5^3 B_6^3 B_3^8+2414513024 B_1^3 B_2^5 B_4^5 B_5^3 B_6^3 B_3^8+766975440 B_1^3 B_2^5 B_4^6 B_5^2
B_6^3 B_3^8+871627680 B_1^3 B_2^6 B_4^5 B_5^2 B_6^3 B_3^8+887409600 B_1^2 B_2^5 B_4^5 B_5^3 B_6^5 B_3^7+887409600 B_1^3 B_2^5 B_4^5 B_5^2 B_6^5 B_3^7+50820000
B_1^2 B_2^4 B_4^6 B_5^4 B_6^4 B_3^7+99607200 B_1^2 B_2^5 B_4^5 B_5^4 B_6^4 B_3^7+3252073440 B_1^2 B_2^5 B_4^6 B_5^3 B_6^4 B_3^7+622545000 B_1^3 B_2^4
B_4^6 B_5^3 B_6^4 B_3^7+2075150000 B_1^2 B_2^6 B_4^5 B_5^3 B_6^4 B_3^7+19480302576 B_1^3 B_2^5 B_4^5 B_5^3 B_6^4 B_3^7+622545000 B_1^3 B_2^6 B_4^4
B_5^3 B_6^4 B_3^7+2075150000 B_1^3 B_2^5 B_4^6 B_5^2 B_6^4 B_3^7+3252073440 B_1^3 B_2^6 B_4^5 B_5^2 B_6^4 B_3^7+99607200 B_1^4 B_2^5 B_4^5 B_5^2
B_6^4 B_3^7+50820000 B_1^4 B_2^6 B_4^4 B_5^2 B_6^4 B_3^7+73045280 B_1^2 B_2^5 B_4^6 B_5^4 B_6^3 B_3^7+21913584 B_1^3 B_2^5 B_4^5 B_5^4 B_6^3 B_3^7+1016898960
B_1^3 B_2^5 B_4^6 B_5^3 B_6^3 B_3^7+1016898960 B_1^3 B_2^6 B_4^5 B_5^3 B_6^3 B_3^7+21913584 B_1^4 B_2^5 B_4^5 B_5^3 B_6^3 B_3^7+73045280 B_1^4 B_2^6
B_4^5 B_5^2 B_6^3 B_3^7) \textbf{Z}^{27}+(356548500 B_1^2 B_2^4 B_4^5 B_5^2 B_6^5 B_3^8+356548500 B_1^2 B_2^5 B_4^4 B_5^2 B_6^5 B_3^8+48024900
B_1 B_2^4 B_4^6 B_5^3 B_6^4 B_3^8+94128804 B_1 B_2^5 B_4^5 B_5^3 B_6^4 B_3^8+5042614500 B_1^2 B_2^4 B_4^5 B_5^3 B_6^4 B_3^8+1961016750 B_1^2 B_2^5
B_4^4 B_5^3 B_6^4 B_3^8+588305025 B_1^2 B_2^4 B_4^6 B_5^2 B_6^4 B_3^8+27835512516 B_1^2 B_2^5 B_4^5 B_5^2 B_6^4 B_3^8+1961016750 B_1^3 B_2^4 B_4^5
B_5^2 B_6^4 B_3^8+588305025 B_1^2 B_2^6 B_4^4 B_5^2 B_6^4 B_3^8+5042614500 B_1^3 B_2^5 B_4^4 B_5^2 B_6^4 B_3^8+94128804 B_1^3 B_2^5 B_4^5 B_5 B_6^4
B_3^8+48024900 B_1^3 B_2^6 B_4^4 B_5 B_6^4 B_3^8+264136950 B_1^2 B_2^4 B_4^6 B_5^3 B_6^3 B_3^8+4790284884 B_1^2 B_2^5 B_4^5 B_5^3 B_6^3 B_3^8+880456500
B_1^2 B_2^5 B_4^6 B_5^2 B_6^3 B_3^8+880456500 B_1^2 B_2^6 B_4^5 B_5^2 B_6^3 B_3^8+4790284884 B_1^3 B_2^5 B_4^5 B_5^2 B_6^3 B_3^8+264136950 B_1^3
B_2^6 B_4^4 B_5^2 B_6^3 B_3^8+305613000 B_1^2 B_2^4 B_4^5 B_5^3 B_6^5 B_3^7+1669054464 B_1^2 B_2^5 B_4^5 B_5^2 B_6^5 B_3^7+305613000 B_1^3 B_2^5
B_4^4 B_5^2 B_6^5 B_3^7+34303500 B_1^2 B_2^4 B_4^5 B_5^4 B_6^4 B_3^7+1413304200 B_1^2 B_2^4 B_4^6 B_5^3 B_6^4 B_3^7+30824064054 B_1^2 B_2^5 B_4^5
B_5^3 B_6^4 B_3^7+6565308750 B_1^3 B_2^4 B_4^5 B_5^3 B_6^4 B_3^7+6565308750 B_1^3 B_2^5 B_4^4 B_5^3 B_6^4 B_3^7+3902976000 B_1^2 B_2^5 B_4^6 B_5^2
B_6^4 B_3^7+3902976000 B_1^2 B_2^6 B_4^5 B_5^2 B_6^4 B_3^7+30824064054 B_1^3 B_2^5 B_4^5 B_5^2 B_6^4 B_3^7+1413304200 B_1^3 B_2^6 B_4^4 B_5^2 B_6^4
B_3^7+34303500 B_1^4 B_2^5 B_4^4 B_5^2 B_6^4 B_3^7+25155900 B_1^2 B_2^4 B_4^6 B_5^4 B_6^3 B_3^7+49305564 B_1^2 B_2^5 B_4^5 B_5^4 B_6^3 B_3^7+1445944500
B_1^2 B_2^5 B_4^6 B_5^3 B_6^3 B_3^7+308159775 B_1^3 B_2^4 B_4^6 B_5^3 B_6^3 B_3^7+1027199250 B_1^2 B_2^6 B_4^5 B_5^3 B_6^3 B_3^7+8951787306 B_1^3
B_2^5 B_4^5 B_5^3 B_6^3 B_3^7+308159775 B_1^3 B_2^6 B_4^4 B_5^3 B_6^3 B_3^7+1027199250 B_1^3 B_2^5 B_4^6 B_5^2 B_6^3 B_3^7+1445944500 B_1^3 B_2^6
B_4^5 B_5^2 B_6^3 B_3^7+49305564 B_1^4 B_2^5 B_4^5 B_5^2 B_6^3 B_3^7+25155900 B_1^4 B_2^6 B_4^4 B_5^2 B_6^3 B_3^7+8199664704 B_1^3 B_2^5 B_4^5 B_5^3
B_6^4 B_3^6) \textbf{Z}^{26}+(409812480 B_1 B_2^4 B_4^5 B_5^3 B_6^4 B_3^8+122943744 B_1 B_2^5 B_4^5 B_5^2 B_6^4 B_3^8+7991343360 B_1^2 B_2^4 B_4^5
B_5^2 B_6^4 B_3^8+7991343360 B_1^2 B_2^5 B_4^4 B_5^2 B_6^4 B_3^8+122943744 B_1^2 B_2^5 B_4^5 B_5 B_6^4 B_3^8+409812480 B_1^3 B_2^5 B_4^4 B_5 B_6^4
B_3^8+2253968640 B_1^2 B_2^4 B_4^5 B_5^3 B_6^3 B_3^8+8611029504 B_1^2 B_2^5 B_4^5 B_5^2 B_6^3 B_3^8+2253968640 B_1^3 B_2^5 B_4^4 B_5^2 B_6^3 B_3^8+599001480
B_1^2 B_2^4 B_4^5 B_5^2 B_6^5 B_3^7+599001480 B_1^2 B_2^5 B_4^4 B_5^2 B_6^5 B_3^7+114345000 B_1 B_2^4 B_4^6 B_5^3 B_6^4 B_3^7+224116200 B_1 B_2^5
B_4^5 B_5^3 B_6^4 B_3^7+19609868880 B_1^2 B_2^4 B_4^5 B_5^3 B_6^4 B_3^7+9158882040 B_1^2 B_2^5 B_4^4 B_5^3 B_6^4 B_3^7+2858625000 B_1^3 B_2^4 B_4^4
B_5^3 B_6^4 B_3^7+1400726250 B_1^2 B_2^4 B_4^6 B_5^2 B_6^4 B_3^7+59798117736 B_1^2 B_2^5 B_4^5 B_5^2 B_6^4 B_3^7+9158882040 B_1^3 B_2^4 B_4^5 B_5^2
B_6^4 B_3^7+1400726250 B_1^2 B_2^6 B_4^4 B_5^2 B_6^4 B_3^7+19609868880 B_1^3 B_2^5 B_4^4 B_5^2 B_6^4 B_3^7+224116200 B_1^3 B_2^5 B_4^5 B_5 B_6^4
B_3^7+114345000 B_1^3 B_2^6 B_4^4 B_5 B_6^4 B_3^7+30187080 B_1^2 B_2^4 B_4^5 B_5^4 B_6^3 B_3^7+966735000 B_1^2 B_2^4 B_4^6 B_5^3 B_6^3 B_3^7+17946116496
B_1^2 B_2^5 B_4^5 B_5^3 B_6^3 B_3^7+3421632060 B_1^3 B_2^4 B_4^5 B_5^3 B_6^3 B_3^7+3421632060 B_1^3 B_2^5 B_4^4 B_5^3 B_6^3 B_3^7+2096325000 B_1^2
B_2^5 B_4^6 B_5^2 B_6^3 B_3^7+2096325000 B_1^2 B_2^6 B_4^5 B_5^2 B_6^3 B_3^7+17946116496 B_1^3 B_2^5 B_4^5 B_5^2 B_6^3 B_3^7+966735000 B_1^3 B_2^6
B_4^4 B_5^2 B_6^3 B_3^7+30187080 B_1^4 B_2^5 B_4^4 B_5^2 B_6^3 B_3^7+439267752 B_1^3 B_2^5 B_4^5 B_5^3 B_6^2 B_3^7+16734009600 B_1^2 B_2^5 B_4^5
B_5^3 B_6^4 B_3^6+5020202880 B_1^3 B_2^4 B_4^5 B_5^3 B_6^4 B_3^6+5020202880 B_1^3 B_2^5 B_4^4 B_5^3 B_6^4 B_3^6+16734009600 B_1^3 B_2^5 B_4^5 B_5^2
B_6^4 B_3^6+6754454784 B_1^3 B_2^5 B_4^5 B_5^3 B_6^3 B_3^6) \textbf{Z}^{25}+(557800320 B_1 B_2^4 B_4^5 B_5^2 B_6^4 B_3^8+1859334400 B_1^2 B_2^4
B_4^4 B_5^2 B_6^4 B_3^8+557800320 B_1^2 B_2^5 B_4^4 B_5 B_6^4 B_3^8+3067901760 B_1^2 B_2^4 B_4^5 B_5^2 B_6^3 B_3^8+3067901760 B_1^2 B_2^5 B_4^4 B_5^2
B_6^3 B_3^8+221852400 B_1^2 B_2^4 B_4^4 B_5^2 B_6^5 B_3^7+2212838320 B_1 B_2^4 B_4^5 B_5^3 B_6^4 B_3^7+1985156250 B_1^2 B_2^3 B_4^5 B_5^3 B_6^4 B_3^7+6097637700
B_1^2 B_2^4 B_4^4 B_5^3 B_6^4 B_3^7+520396800 B_1 B_2^5 B_4^5 B_5^2 B_6^4 B_3^7+40830215370 B_1^2 B_2^4 B_4^5 B_5^2 B_6^4 B_3^7+40830215370 B_1^2
B_2^5 B_4^4 B_5^2 B_6^4 B_3^7+6097637700 B_1^3 B_2^4 B_4^4 B_5^2 B_6^4 B_3^7+1985156250 B_1^3 B_2^5 B_4^3 B_5^2 B_6^4 B_3^7+520396800 B_1^2 B_2^5
B_4^5 B_5 B_6^4 B_3^7+2212838320 B_1^3 B_2^5 B_4^4 B_5 B_6^4 B_3^7+69877500 B_1 B_2^4 B_4^6 B_5^3 B_6^3 B_3^7+136959900 B_1 B_2^5 B_4^5 B_5^3 B_6^3
B_3^7+16980581310 B_1^2 B_2^4 B_4^5 B_5^3 B_6^3 B_3^7+5281747240 B_1^2 B_2^5 B_4^4 B_5^3 B_6^3 B_3^7+2862182400 B_1^3 B_2^4 B_4^4 B_5^3 B_6^3 B_3^7+855999375
B_1^2 B_2^4 B_4^6 B_5^2 B_6^3 B_3^7+41763159900 B_1^2 B_2^5 B_4^5 B_5^2 B_6^3 B_3^7+5281747240 B_1^3 B_2^4 B_4^5 B_5^2 B_6^3 B_3^7+855999375 B_1^2
B_2^6 B_4^4 B_5^2 B_6^3 B_3^7+16980581310 B_1^3 B_2^5 B_4^4 B_5^2 B_6^3 B_3^7+136959900 B_1^3 B_2^5 B_4^5 B_5 B_6^3 B_3^7+69877500 B_1^3 B_2^6 B_4^4
B_5 B_6^3 B_3^7+1220188200 B_1^2 B_2^5 B_4^5 B_5^3 B_6^2 B_3^7+1220188200 B_1^3 B_2^5 B_4^5 B_5^2 B_6^2 B_3^7+18993314340 B_1^2 B_2^4 B_4^5 B_5^3
B_6^4 B_3^6+10676646750 B_1^2 B_2^5 B_4^4 B_5^3 B_6^4 B_3^6+3890016900 B_1^3 B_2^4 B_4^4 B_5^3 B_6^4 B_3^6+38856294400 B_1^2 B_2^5 B_4^5 B_5^2 B_6^4
B_3^6+10676646750 B_1^3 B_2^4 B_4^5 B_5^2 B_6^4 B_3^6+18993314340 B_1^3 B_2^5 B_4^4 B_5^2 B_6^4 B_3^6+3913140 B_1^2 B_2^4 B_4^5 B_5^4 B_6^3 B_3^6+81523750
B_1^2 B_2^4 B_4^6 B_5^3 B_6^3 B_3^6+16798204500 B_1^2 B_2^5 B_4^5 B_5^3 B_6^3 B_3^6+5039461350 B_1^3 B_2^4 B_4^5 B_5^3 B_6^3 B_3^6+5039461350 B_1^3
B_2^5 B_4^4 B_5^3 B_6^3 B_3^6+16798204500 B_1^3 B_2^5 B_4^5 B_5^2 B_6^3 B_3^6+81523750 B_1^3 B_2^6 B_4^4 B_5^2 B_6^3 B_3^6+3913140 B_1^4 B_2^5 B_4^4
B_5^2 B_6^3 B_3^6+1423552900 B_1^3 B_2^5 B_4^5 B_5^3 B_6^2 B_3^6) \textbf{Z}^{24}+(457380000 B_1 B_2^3 B_4^5 B_5^3 B_6^4 B_3^7+896464800 B_1 B_2^4
B_4^4 B_5^3 B_6^4 B_3^7+4201797600 B_1 B_2^4 B_4^5 B_5^2 B_6^4 B_3^7+5602905000 B_1^2 B_2^3 B_4^5 B_5^2 B_6^4 B_3^7+268939440 B_1 B_2^5 B_4^4 B_5^2
B_6^4 B_3^7+32647658400 B_1^2 B_2^4 B_4^4 B_5^2 B_6^4 B_3^7+5602905000 B_1^2 B_2^5 B_4^3 B_5^2 B_6^4 B_3^7+268939440 B_1^2 B_2^4 B_4^5 B_5 B_6^4
B_3^7+4201797600 B_1^2 B_2^5 B_4^4 B_5 B_6^4 B_3^7+896464800 B_1^3 B_2^4 B_4^4 B_5 B_6^4 B_3^7+457380000 B_1^3 B_2^5 B_4^3 B_5 B_6^4 B_3^7+1537683840
B_1 B_2^4 B_4^5 B_5^3 B_6^3 B_3^7+2515590000 B_1^2 B_2^3 B_4^5 B_5^3 B_6^3 B_3^7+6940533600 B_1^2 B_2^4 B_4^4 B_5^3 B_6^3 B_3^7+402494400 B_1 B_2^5
B_4^5 B_5^2 B_6^3 B_3^7+38328942960 B_1^2 B_2^4 B_4^5 B_5^2 B_6^3 B_3^7+38328942960 B_1^2 B_2^5 B_4^4 B_5^2 B_6^3 B_3^7+6940533600 B_1^3 B_2^4 B_4^4
B_5^2 B_6^3 B_3^7+2515590000 B_1^3 B_2^5 B_4^3 B_5^2 B_6^3 B_3^7+402494400 B_1^2 B_2^5 B_4^5 B_5 B_6^3 B_3^7+1537683840 B_1^3 B_2^5 B_4^4 B_5 B_6^3
B_3^7+1075757760 B_1^2 B_2^4 B_4^5 B_5^3 B_6^2 B_3^7+3585859200 B_1^2 B_2^5 B_4^5 B_5^2 B_6^2 B_3^7+1075757760 B_1^3 B_2^5 B_4^4 B_5^2 B_6^2 B_3^7+2561328000
B_1 B_2^4 B_4^5 B_5^3 B_6^4 B_3^6+4802490000 B_1^2 B_2^3 B_4^5 B_5^3 B_6^4 B_3^6+14386125600 B_1^2 B_2^4 B_4^4 B_5^3 B_6^4 B_3^6+48870138240 B_1^2
B_2^4 B_4^5 B_5^2 B_6^4 B_3^6+48870138240 B_1^2 B_2^5 B_4^4 B_5^2 B_6^4 B_3^6+14386125600 B_1^3 B_2^4 B_4^4 B_5^2 B_6^4 B_3^6+4802490000 B_1^3 B_2^5
B_4^3 B_5^2 B_6^4 B_3^6+2561328000 B_1^3 B_2^5 B_4^4 B_5 B_6^4 B_3^6+28131531120 B_1^2 B_2^4 B_4^5 B_5^3 B_6^3 B_3^6+12664602720 B_1^2 B_2^5 B_4^4
B_5^3 B_6^3 B_3^6+6411081600 B_1^3 B_2^4 B_4^4 B_5^3 B_6^3 B_3^6+45964195200 B_1^2 B_2^5 B_4^5 B_5^2 B_6^3 B_3^6+12664602720 B_1^3 B_2^4 B_4^5 B_5^2
B_6^3 B_3^6+28131531120 B_1^3 B_2^5 B_4^4 B_5^2 B_6^3 B_3^6+4183502400 B_1^2 B_2^5 B_4^5 B_5^3 B_6^2 B_3^6+1255050720 B_1^3 B_2^4 B_4^5 B_5^3 B_6^2
B_3^6+1255050720 B_1^3 B_2^5 B_4^4 B_5^3 B_6^2 B_3^6+4183502400 B_1^3 B_2^5 B_4^5 B_5^2 B_6^2 B_3^6) \textbf{Z}^{23}+(1400726250 B_1 B_2^3 B_4^5
B_5^2 B_6^4 B_3^7+3186932364 B_1 B_2^4 B_4^4 B_5^2 B_6^4 B_3^7+4669087500 B_1^2 B_2^3 B_4^4 B_5^2 B_6^4 B_3^7+4669087500 B_1^2 B_2^4 B_4^3 B_5^2
B_6^4 B_3^7+3186932364 B_1^2 B_2^4 B_4^4 B_5 B_6^4 B_3^7+1400726250 B_1^2 B_2^5 B_4^3 B_5 B_6^4 B_3^7+628897500 B_1 B_2^3 B_4^5 B_5^3 B_6^3 B_3^7+1232639100
B_1 B_2^4 B_4^4 B_5^3 B_6^3 B_3^7+3421632060 B_1 B_2^4 B_4^5 B_5^2 B_6^3 B_3^7+7703994375 B_1^2 B_2^3 B_4^5 B_5^2 B_6^3 B_3^7+369791730 B_1 B_2^5
B_4^4 B_5^2 B_6^3 B_3^7+38630489436 B_1^2 B_2^4 B_4^4 B_5^2 B_6^3 B_3^7+7703994375 B_1^2 B_2^5 B_4^3 B_5^2 B_6^3 B_3^7+369791730 B_1^2 B_2^4 B_4^5
B_5 B_6^3 B_3^7+3421632060 B_1^2 B_2^5 B_4^4 B_5 B_6^3 B_3^7+1232639100 B_1^3 B_2^4 B_4^4 B_5 B_6^3 B_3^7+628897500 B_1^3 B_2^5 B_4^3 B_5 B_6^3 B_3^7+3294508140
B_1^2 B_2^4 B_4^5 B_5^2 B_6^2 B_3^7+3294508140 B_1^2 B_2^5 B_4^4 B_5^2 B_6^2 B_3^7+1200622500 B_1 B_2^3 B_4^5 B_5^3 B_6^4 B_3^6+2353220100 B_1 B_2^4
B_4^4 B_5^3 B_6^4 B_3^6+4002075000 B_1^2 B_2^3 B_4^4 B_5^3 B_6^4 B_3^6+6556999680 B_1 B_2^4 B_4^5 B_5^2 B_6^4 B_3^6+14707625625 B_1^2 B_2^3 B_4^5
B_5^2 B_6^4 B_3^6+76881768516 B_1^2 B_2^4 B_4^4 B_5^2 B_6^4 B_3^6+14707625625 B_1^2 B_2^5 B_4^3 B_5^2 B_6^4 B_3^6+4002075000 B_1^3 B_2^4 B_4^3 B_5^2
B_6^4 B_3^6+6556999680 B_1^2 B_2^5 B_4^4 B_5 B_6^4 B_3^6+2353220100 B_1^3 B_2^4 B_4^4 B_5 B_6^4 B_3^6+1200622500 B_1^3 B_2^5 B_4^3 B_5 B_6^4 B_3^6+3447314640
B_1 B_2^4 B_4^5 B_5^3 B_6^3 B_3^6+10017315000 B_1^2 B_2^3 B_4^5 B_5^3 B_6^3 B_3^6+27874845306 B_1^2 B_2^4 B_4^4 B_5^3 B_6^3 B_3^6+80030488230 B_1^2
B_2^4 B_4^5 B_5^2 B_6^3 B_3^6+80030488230 B_1^2 B_2^5 B_4^4 B_5^2 B_6^3 B_3^6+27874845306 B_1^3 B_2^4 B_4^4 B_5^2 B_6^3 B_3^6+10017315000 B_1^3 B_2^5
B_4^3 B_5^2 B_6^3 B_3^6+3447314640 B_1^3 B_2^5 B_4^4 B_5 B_6^3 B_3^6+8519617260 B_1^2 B_2^4 B_4^5 B_5^3 B_6^2 B_3^6+3843592830 B_1^2 B_2^5 B_4^4
B_5^3 B_6^2 B_3^6+1967099904 B_1^3 B_2^4 B_4^4 B_5^3 B_6^2 B_3^6+13340250000 B_1^2 B_2^5 B_4^5 B_5^2 B_6^2 B_3^6+3843592830 B_1^3 B_2^4 B_4^5 B_5^2
B_6^2 B_3^6+8519617260 B_1^3 B_2^5 B_4^4 B_5^2 B_6^2 B_3^6+2629630080 B_1^2 B_2^4 B_4^5 B_5^3 B_6^3 B_3^5+788889024 B_1^3 B_2^4 B_4^4 B_5^3 B_6^3
B_3^5+2629630080 B_1^3 B_2^5 B_4^4 B_5^2 B_6^3 B_3^5) \textbf{Z}^{22}+(1328096000 B_1 B_2^3 B_4^4 B_5^2 B_6^4 B_3^7+398428800 B_1 B_2^4 B_4^4 B_5
B_6^4 B_3^7+1328096000 B_1^2 B_2^4 B_4^3 B_5 B_6^4 B_3^7+2191358400 B_1 B_2^3 B_4^5 B_5^2 B_6^3 B_3^7+4881115008 B_1 B_2^4 B_4^4 B_5^2 B_6^3 B_3^7+7304528000
B_1^2 B_2^3 B_4^4 B_5^2 B_6^3 B_3^7+7304528000 B_1^2 B_2^4 B_4^3 B_5^2 B_6^3 B_3^7+4881115008 B_1^2 B_2^4 B_4^4 B_5 B_6^3 B_3^7+2191358400 B_1^2
B_2^5 B_4^3 B_5 B_6^3 B_3^7+3123681792 B_1^2 B_2^4 B_4^4 B_5^2 B_6^2 B_3^7+1138368000 B_1 B_2^3 B_4^4 B_5^3 B_6^4 B_3^6+3890016900 B_1 B_2^3 B_4^5
B_5^2 B_6^4 B_3^6+10875161528 B_1 B_2^4 B_4^4 B_5^2 B_6^4 B_3^6+28921662000 B_1^2 B_2^3 B_4^4 B_5^2 B_6^4 B_3^6+28921662000 B_1^2 B_2^4 B_4^3 B_5^2
B_6^4 B_3^6+10875161528 B_1^2 B_2^4 B_4^4 B_5 B_6^4 B_3^6+3890016900 B_1^2 B_2^5 B_4^3 B_5 B_6^4 B_3^6+1138368000 B_1^3 B_2^4 B_4^3 B_5 B_6^4 B_3^6+2752521200
B_1 B_2^3 B_4^5 B_5^3 B_6^3 B_3^6+5394941552 B_1 B_2^4 B_4^4 B_5^3 B_6^3 B_3^6+10284615000 B_1^2 B_2^3 B_4^4 B_5^3 B_6^3 B_3^6+97828500 B_1^2 B_2^4
B_4^3 B_5^3 B_6^3 B_3^6+10284615000 B_1 B_2^4 B_4^5 B_5^2 B_6^3 B_3^6+33718384700 B_1^2 B_2^3 B_4^5 B_5^2 B_6^3 B_3^6+97828500 B_1 B_2^5 B_4^4 B_5^2
B_6^3 B_3^6+163830961008 B_1^2 B_2^4 B_4^4 B_5^2 B_6^3 B_3^6+97828500 B_1^3 B_2^3 B_4^4 B_5^2 B_6^3 B_3^6+33718384700 B_1^2 B_2^5 B_4^3 B_5^2 B_6^3
B_3^6+10284615000 B_1^3 B_2^4 B_4^3 B_5^2 B_6^3 B_3^6+97828500 B_1^2 B_2^4 B_4^5 B_5 B_6^3 B_3^6+10284615000 B_1^2 B_2^5 B_4^4 B_5 B_6^3 B_3^6+5394941552
B_1^3 B_2^4 B_4^4 B_5 B_6^3 B_3^6+2752521200 B_1^3 B_2^5 B_4^3 B_5 B_6^3 B_3^6+1138368000 B_1 B_2^4 B_4^5 B_5^3 B_6^2 B_3^6+3890016900 B_1^2 B_2^3
B_4^5 B_5^3 B_6^2 B_3^6+10875161528 B_1^2 B_2^4 B_4^4 B_5^3 B_6^2 B_3^6+28921662000 B_1^2 B_2^4 B_4^5 B_5^2 B_6^2 B_3^6+28921662000 B_1^2 B_2^5 B_4^4
B_5^2 B_6^2 B_3^6+10875161528 B_1^3 B_2^4 B_4^4 B_5^2 B_6^2 B_3^6+3890016900 B_1^3 B_2^5 B_4^3 B_5^2 B_6^2 B_3^6+1138368000 B_1^3 B_2^5 B_4^4 B_5
B_6^2 B_3^6+3123681792 B_1^2 B_2^4 B_4^4 B_5^2 B_6^4 B_3^5+2191358400 B_1^2 B_2^3 B_4^5 B_5^3 B_6^3 B_3^5+4881115008 B_1^2 B_2^4 B_4^4 B_5^3 B_6^3
B_3^5+7304528000 B_1^2 B_2^4 B_4^5 B_5^2 B_6^3 B_3^5+7304528000 B_1^2 B_2^5 B_4^4 B_5^2 B_6^3 B_3^5+4881115008 B_1^3 B_2^4 B_4^4 B_5^2 B_6^3 B_3^5+2191358400
B_1^3 B_2^5 B_4^3 B_5^2 B_6^3 B_3^5+1328096000 B_1^2 B_2^4 B_4^5 B_5^3 B_6^2 B_3^5+398428800 B_1^3 B_2^4 B_4^4 B_5^3 B_6^2 B_3^5+1328096000 B_1^3
B_2^5 B_4^4 B_5^2 B_6^2 B_3^5) \textbf{Z}^{21}+(2629630080 B_1 B_2^3 B_4^4 B_5^2 B_6^3 B_3^7+788889024 B_1 B_2^4 B_4^4 B_5 B_6^3 B_3^7+2629630080
B_1^2 B_2^4 B_4^3 B_5 B_6^3 B_3^7+8519617260 B_1 B_2^3 B_4^4 B_5^2 B_6^4 B_3^6+3843592830 B_1 B_2^4 B_4^3 B_5^2 B_6^4 B_3^6+13340250000 B_1^2 B_2^3
B_4^3 B_5^2 B_6^4 B_3^6+1967099904 B_1 B_2^4 B_4^4 B_5 B_6^4 B_3^6+3843592830 B_1^2 B_2^3 B_4^4 B_5 B_6^4 B_3^6+8519617260 B_1^2 B_2^4 B_4^3 B_5
B_6^4 B_3^6+3447314640 B_1 B_2^3 B_4^4 B_5^3 B_6^3 B_3^6+10017315000 B_1 B_2^3 B_4^5 B_5^2 B_6^3 B_3^6+27874845306 B_1 B_2^4 B_4^4 B_5^2 B_6^3 B_3^6+80030488230
B_1^2 B_2^3 B_4^4 B_5^2 B_6^3 B_3^6+80030488230 B_1^2 B_2^4 B_4^3 B_5^2 B_6^3 B_3^6+27874845306 B_1^2 B_2^4 B_4^4 B_5 B_6^3 B_3^6+10017315000 B_1^2
B_2^5 B_4^3 B_5 B_6^3 B_3^6+3447314640 B_1^3 B_2^4 B_4^3 B_5 B_6^3 B_3^6+1200622500 B_1 B_2^3 B_4^5 B_5^3 B_6^2 B_3^6+2353220100 B_1 B_2^4 B_4^4
B_5^3 B_6^2 B_3^6+6556999680 B_1^2 B_2^3 B_4^4 B_5^3 B_6^2 B_3^6+4002075000 B_1 B_2^4 B_4^5 B_5^2 B_6^2 B_3^6+14707625625 B_1^2 B_2^3 B_4^5 B_5^2
B_6^2 B_3^6+76881768516 B_1^2 B_2^4 B_4^4 B_5^2 B_6^2 B_3^6+14707625625 B_1^2 B_2^5 B_4^3 B_5^2 B_6^2 B_3^6+6556999680 B_1^3 B_2^4 B_4^3 B_5^2 B_6^2
B_3^6+4002075000 B_1^2 B_2^5 B_4^4 B_5 B_6^2 B_3^6+2353220100 B_1^3 B_2^4 B_4^4 B_5 B_6^2 B_3^6+1200622500 B_1^3 B_2^5 B_4^3 B_5 B_6^2 B_3^6+3294508140
B_1^2 B_2^3 B_4^4 B_5^2 B_6^4 B_3^5+3294508140 B_1^2 B_2^4 B_4^3 B_5^2 B_6^4 B_3^5+628897500 B_1 B_2^3 B_4^5 B_5^3 B_6^3 B_3^5+1232639100 B_1 B_2^4
B_4^4 B_5^3 B_6^3 B_3^5+3421632060 B_1^2 B_2^3 B_4^4 B_5^3 B_6^3 B_3^5+369791730 B_1^2 B_2^4 B_4^3 B_5^3 B_6^3 B_3^5+7703994375 B_1^2 B_2^3 B_4^5
B_5^2 B_6^3 B_3^5+38630489436 B_1^2 B_2^4 B_4^4 B_5^2 B_6^3 B_3^5+369791730 B_1^3 B_2^3 B_4^4 B_5^2 B_6^3 B_3^5+7703994375 B_1^2 B_2^5 B_4^3 B_5^2
B_6^3 B_3^5+3421632060 B_1^3 B_2^4 B_4^3 B_5^2 B_6^3 B_3^5+1232639100 B_1^3 B_2^4 B_4^4 B_5 B_6^3 B_3^5+628897500 B_1^3 B_2^5 B_4^3 B_5 B_6^3 B_3^5+1400726250
B_1^2 B_2^3 B_4^5 B_5^3 B_6^2 B_3^5+3186932364 B_1^2 B_2^4 B_4^4 B_5^3 B_6^2 B_3^5+4669087500 B_1^2 B_2^4 B_4^5 B_5^2 B_6^2 B_3^5+4669087500 B_1^2
B_2^5 B_4^4 B_5^2 B_6^2 B_3^5+3186932364 B_1^3 B_2^4 B_4^4 B_5^2 B_6^2 B_3^5+1400726250 B_1^3 B_2^5 B_4^3 B_5^2 B_6^2 B_3^5) \textbf{Z}^{20}+(4183502400
B_1 B_2^3 B_4^3 B_5^2 B_6^4 B_3^6+1255050720 B_1 B_2^3 B_4^4 B_5 B_6^4 B_3^6+1255050720 B_1 B_2^4 B_4^3 B_5 B_6^4 B_3^6+4183502400 B_1^2 B_2^3 B_4^3
B_5 B_6^4 B_3^6+28131531120 B_1 B_2^3 B_4^4 B_5^2 B_6^3 B_3^6+12664602720 B_1 B_2^4 B_4^3 B_5^2 B_6^3 B_3^6+45964195200 B_1^2 B_2^3 B_4^3 B_5^2 B_6^3
B_3^6+6411081600 B_1 B_2^4 B_4^4 B_5 B_6^3 B_3^6+12664602720 B_1^2 B_2^3 B_4^4 B_5 B_6^3 B_3^6+28131531120 B_1^2 B_2^4 B_4^3 B_5 B_6^3 B_3^6+2561328000
B_1 B_2^3 B_4^4 B_5^3 B_6^2 B_3^6+4802490000 B_1 B_2^3 B_4^5 B_5^2 B_6^2 B_3^6+14386125600 B_1 B_2^4 B_4^4 B_5^2 B_6^2 B_3^6+48870138240 B_1^2 B_2^3
B_4^4 B_5^2 B_6^2 B_3^6+48870138240 B_1^2 B_2^4 B_4^3 B_5^2 B_6^2 B_3^6+14386125600 B_1^2 B_2^4 B_4^4 B_5 B_6^2 B_3^6+4802490000 B_1^2 B_2^5 B_4^3
B_5 B_6^2 B_3^6+2561328000 B_1^3 B_2^4 B_4^3 B_5 B_6^2 B_3^6+1075757760 B_1 B_2^3 B_4^4 B_5^2 B_6^4 B_3^5+3585859200 B_1^2 B_2^3 B_4^3 B_5^2 B_6^4
B_3^5+1075757760 B_1^2 B_2^4 B_4^3 B_5 B_6^4 B_3^5+1537683840 B_1 B_2^3 B_4^4 B_5^3 B_6^3 B_3^5+402494400 B_1^2 B_2^3 B_4^3 B_5^3 B_6^3 B_3^5+2515590000
B_1 B_2^3 B_4^5 B_5^2 B_6^3 B_3^5+6940533600 B_1 B_2^4 B_4^4 B_5^2 B_6^3 B_3^5+38328942960 B_1^2 B_2^3 B_4^4 B_5^2 B_6^3 B_3^5+38328942960 B_1^2
B_2^4 B_4^3 B_5^2 B_6^3 B_3^5+402494400 B_1^3 B_2^3 B_4^3 B_5^2 B_6^3 B_3^5+6940533600 B_1^2 B_2^4 B_4^4 B_5 B_6^3 B_3^5+2515590000 B_1^2 B_2^5 B_4^3
B_5 B_6^3 B_3^5+1537683840 B_1^3 B_2^4 B_4^3 B_5 B_6^3 B_3^5+457380000 B_1 B_2^3 B_4^5 B_5^3 B_6^2 B_3^5+896464800 B_1 B_2^4 B_4^4 B_5^3 B_6^2 B_3^5+4201797600
B_1^2 B_2^3 B_4^4 B_5^3 B_6^2 B_3^5+268939440 B_1^2 B_2^4 B_4^3 B_5^3 B_6^2 B_3^5+5602905000 B_1^2 B_2^3 B_4^5 B_5^2 B_6^2 B_3^5+32647658400 B_1^2
B_2^4 B_4^4 B_5^2 B_6^2 B_3^5+268939440 B_1^3 B_2^3 B_4^4 B_5^2 B_6^2 B_3^5+5602905000 B_1^2 B_2^5 B_4^3 B_5^2 B_6^2 B_3^5+4201797600 B_1^3 B_2^4
B_4^3 B_5^2 B_6^2 B_3^5+896464800 B_1^3 B_2^4 B_4^4 B_5 B_6^2 B_3^5+457380000 B_1^3 B_2^5 B_4^3 B_5 B_6^2 B_3^5) \textbf{Z}^{19}+(1423552900 B_1
B_2^3 B_4^3 B_5 B_6^4 B_3^6+3913140 B_1^2 B_2^4 B_4^3 B_6^3 B_3^6+3913140 B_2^3 B_4^4 B_5^2 B_6^3 B_3^6+81523750 B_1 B_2^2 B_4^4 B_5^2 B_6^3 B_3^6+16798204500
B_1 B_2^3 B_4^3 B_5^2 B_6^3 B_3^6+5039461350 B_1 B_2^3 B_4^4 B_5 B_6^3 B_3^6+5039461350 B_1 B_2^4 B_4^3 B_5 B_6^3 B_3^6+16798204500 B_1^2 B_2^3 B_4^3
B_5 B_6^3 B_3^6+81523750 B_1^2 B_2^4 B_4^2 B_5 B_6^3 B_3^6+18993314340 B_1 B_2^3 B_4^4 B_5^2 B_6^2 B_3^6+10676646750 B_1 B_2^4 B_4^3 B_5^2 B_6^2
B_3^6+38856294400 B_1^2 B_2^3 B_4^3 B_5^2 B_6^2 B_3^6+3890016900 B_1 B_2^4 B_4^4 B_5 B_6^2 B_3^6+10676646750 B_1^2 B_2^3 B_4^4 B_5 B_6^2 B_3^6+18993314340
B_1^2 B_2^4 B_4^3 B_5 B_6^2 B_3^6+1220188200 B_1 B_2^3 B_4^3 B_5^2 B_6^4 B_3^5+1220188200 B_1^2 B_2^3 B_4^3 B_5 B_6^4 B_3^5+69877500 B_1 B_2^2 B_4^4
B_5^3 B_6^3 B_3^5+136959900 B_1 B_2^3 B_4^3 B_5^3 B_6^3 B_3^5+16980581310 B_1 B_2^3 B_4^4 B_5^2 B_6^3 B_3^5+855999375 B_1^2 B_2^2 B_4^4 B_5^2 B_6^3
B_3^5+5281747240 B_1 B_2^4 B_4^3 B_5^2 B_6^3 B_3^5+41763159900 B_1^2 B_2^3 B_4^3 B_5^2 B_6^3 B_3^5+855999375 B_1^2 B_2^4 B_4^2 B_5^2 B_6^3 B_3^5+2862182400
B_1 B_2^4 B_4^4 B_5 B_6^3 B_3^5+5281747240 B_1^2 B_2^3 B_4^4 B_5 B_6^3 B_3^5+16980581310 B_1^2 B_2^4 B_4^3 B_5 B_6^3 B_3^5+136959900 B_1^3 B_2^3
B_4^3 B_5 B_6^3 B_3^5+69877500 B_1^3 B_2^4 B_4^2 B_5 B_6^3 B_3^5+2212838320 B_1 B_2^3 B_4^4 B_5^3 B_6^2 B_3^5+520396800 B_1^2 B_2^3 B_4^3 B_5^3 B_6^2
B_3^5+1985156250 B_1 B_2^3 B_4^5 B_5^2 B_6^2 B_3^5+6097637700 B_1 B_2^4 B_4^4 B_5^2 B_6^2 B_3^5+40830215370 B_1^2 B_2^3 B_4^4 B_5^2 B_6^2 B_3^5+40830215370
B_1^2 B_2^4 B_4^3 B_5^2 B_6^2 B_3^5+520396800 B_1^3 B_2^3 B_4^3 B_5^2 B_6^2 B_3^5+6097637700 B_1^2 B_2^4 B_4^4 B_5 B_6^2 B_3^5+1985156250 B_1^2 B_2^5
B_4^3 B_5 B_6^2 B_3^5+2212838320 B_1^3 B_2^4 B_4^3 B_5 B_6^2 B_3^5+221852400 B_1^2 B_2^4 B_4^4 B_5^2 B_6 B_3^5+3067901760 B_1^2 B_2^3 B_4^4 B_5^2
B_6^3 B_3^4+3067901760 B_1^2 B_2^4 B_4^3 B_5^2 B_6^3 B_3^4+557800320 B_1^2 B_2^3 B_4^4 B_5^3 B_6^2 B_3^4+1859334400 B_1^2 B_2^4 B_4^4 B_5^2 B_6^2
B_3^4+557800320 B_1^3 B_2^4 B_4^3 B_5^2 B_6^2 B_3^4) \textbf{Z}^{18}+(6754454784 B_1 B_2^3 B_4^3 B_5 B_6^3 B_3^6+16734009600 B_1 B_2^3 B_4^3 B_5^2
B_6^2 B_3^6+5020202880 B_1 B_2^3 B_4^4 B_5 B_6^2 B_3^6+5020202880 B_1 B_2^4 B_4^3 B_5 B_6^2 B_3^6+16734009600 B_1^2 B_2^3 B_4^3 B_5 B_6^2 B_3^6+439267752
B_1 B_2^3 B_4^3 B_5 B_6^4 B_3^5+30187080 B_1^2 B_2^4 B_4^3 B_6^3 B_3^5+30187080 B_2^3 B_4^4 B_5^2 B_6^3 B_3^5+966735000 B_1 B_2^2 B_4^4 B_5^2 B_6^3
B_3^5+17946116496 B_1 B_2^3 B_4^3 B_5^2 B_6^3 B_3^5+2096325000 B_1^2 B_2^2 B_4^3 B_5^2 B_6^3 B_3^5+2096325000 B_1^2 B_2^3 B_4^2 B_5^2 B_6^3 B_3^5+3421632060
B_1 B_2^3 B_4^4 B_5 B_6^3 B_3^5+3421632060 B_1 B_2^4 B_4^3 B_5 B_6^3 B_3^5+17946116496 B_1^2 B_2^3 B_4^3 B_5 B_6^3 B_3^5+966735000 B_1^2 B_2^4 B_4^2
B_5 B_6^3 B_3^5+114345000 B_1 B_2^2 B_4^4 B_5^3 B_6^2 B_3^5+224116200 B_1 B_2^3 B_4^3 B_5^3 B_6^2 B_3^5+19609868880 B_1 B_2^3 B_4^4 B_5^2 B_6^2 B_3^5+1400726250
B_1^2 B_2^2 B_4^4 B_5^2 B_6^2 B_3^5+9158882040 B_1 B_2^4 B_4^3 B_5^2 B_6^2 B_3^5+59798117736 B_1^2 B_2^3 B_4^3 B_5^2 B_6^2 B_3^5+1400726250 B_1^2
B_2^4 B_4^2 B_5^2 B_6^2 B_3^5+2858625000 B_1 B_2^4 B_4^4 B_5 B_6^2 B_3^5+9158882040 B_1^2 B_2^3 B_4^4 B_5 B_6^2 B_3^5+19609868880 B_1^2 B_2^4 B_4^3
B_5 B_6^2 B_3^5+224116200 B_1^3 B_2^3 B_4^3 B_5 B_6^2 B_3^5+114345000 B_1^3 B_2^4 B_4^2 B_5 B_6^2 B_3^5+599001480 B_1^2 B_2^3 B_4^4 B_5^2 B_6 B_3^5+599001480
B_1^2 B_2^4 B_4^3 B_5^2 B_6 B_3^5+2253968640 B_1 B_2^3 B_4^4 B_5^2 B_6^3 B_3^4+8611029504 B_1^2 B_2^3 B_4^3 B_5^2 B_6^3 B_3^4+2253968640 B_1^2 B_2^4
B_4^3 B_5 B_6^3 B_3^4+409812480 B_1 B_2^3 B_4^4 B_5^3 B_6^2 B_3^4+122943744 B_1^2 B_2^3 B_4^3 B_5^3 B_6^2 B_3^4+7991343360 B_1^2 B_2^3 B_4^4 B_5^2
B_6^2 B_3^4+7991343360 B_1^2 B_2^4 B_4^3 B_5^2 B_6^2 B_3^4+122943744 B_1^3 B_2^3 B_4^3 B_5^2 B_6^2 B_3^4+409812480 B_1^3 B_2^4 B_4^3 B_5 B_6^2 B_3^4)
\textbf{Z}^{17}+(8199664704 B_1 B_2^3 B_4^3 B_5 B_6^2 B_3^6+49305564 B_1^2 B_2^3 B_4^3 B_6^3 B_3^5+25155900 B_1^2 B_2^4 B_4^2 B_6^3 B_3^5+25155900 B_2^2
B_4^4 B_5^2 B_6^3 B_3^5+49305564 B_2^3 B_4^3 B_5^2 B_6^3 B_3^5+1445944500 B_1 B_2^2 B_4^3 B_5^2 B_6^3 B_3^5+1027199250 B_1 B_2^3 B_4^2 B_5^2 B_6^3
B_3^5+308159775 B_1 B_2^2 B_4^4 B_5 B_6^3 B_3^5+8951787306 B_1 B_2^3 B_4^3 B_5 B_6^3 B_3^5+1027199250 B_1^2 B_2^2 B_4^3 B_5 B_6^3 B_3^5+308159775
B_1 B_2^4 B_4^2 B_5 B_6^3 B_3^5+1445944500 B_1^2 B_2^3 B_4^2 B_5 B_6^3 B_3^5+34303500 B_1^2 B_2^4 B_4^3 B_6^2 B_3^5+34303500 B_2^3 B_4^4 B_5^2 B_6^2
B_3^5+1413304200 B_1 B_2^2 B_4^4 B_5^2 B_6^2 B_3^5+30824064054 B_1 B_2^3 B_4^3 B_5^2 B_6^2 B_3^5+3902976000 B_1^2 B_2^2 B_4^3 B_5^2 B_6^2 B_3^5+3902976000
B_1^2 B_2^3 B_4^2 B_5^2 B_6^2 B_3^5+6565308750 B_1 B_2^3 B_4^4 B_5 B_6^2 B_3^5+6565308750 B_1 B_2^4 B_4^3 B_5 B_6^2 B_3^5+30824064054 B_1^2 B_2^3
B_4^3 B_5 B_6^2 B_3^5+1413304200 B_1^2 B_2^4 B_4^2 B_5 B_6^2 B_3^5+305613000 B_1 B_2^3 B_4^4 B_5^2 B_6 B_3^5+1669054464 B_1^2 B_2^3 B_4^3 B_5^2 B_6
B_3^5+305613000 B_1^2 B_2^4 B_4^3 B_5 B_6 B_3^5+264136950 B_1 B_2^2 B_4^4 B_5^2 B_6^3 B_3^4+4790284884 B_1 B_2^3 B_4^3 B_5^2 B_6^3 B_3^4+880456500
B_1^2 B_2^2 B_4^3 B_5^2 B_6^3 B_3^4+880456500 B_1^2 B_2^3 B_4^2 B_5^2 B_6^3 B_3^4+4790284884 B_1^2 B_2^3 B_4^3 B_5 B_6^3 B_3^4+264136950 B_1^2 B_2^4
B_4^2 B_5 B_6^3 B_3^4+48024900 B_1 B_2^2 B_4^4 B_5^3 B_6^2 B_3^4+94128804 B_1 B_2^3 B_4^3 B_5^3 B_6^2 B_3^4+5042614500 B_1 B_2^3 B_4^4 B_5^2 B_6^2
B_3^4+588305025 B_1^2 B_2^2 B_4^4 B_5^2 B_6^2 B_3^4+1961016750 B_1 B_2^4 B_4^3 B_5^2 B_6^2 B_3^4+27835512516 B_1^2 B_2^3 B_4^3 B_5^2 B_6^2 B_3^4+588305025
B_1^2 B_2^4 B_4^2 B_5^2 B_6^2 B_3^4+1961016750 B_1^2 B_2^3 B_4^4 B_5 B_6^2 B_3^4+5042614500 B_1^2 B_2^4 B_4^3 B_5 B_6^2 B_3^4+94128804 B_1^3 B_2^3
B_4^3 B_5 B_6^2 B_3^4+48024900 B_1^3 B_2^4 B_4^2 B_5 B_6^2 B_3^4+356548500 B_1^2 B_2^3 B_4^4 B_5^2 B_6 B_3^4+356548500 B_1^2 B_2^4 B_4^3 B_5^2 B_6
B_3^4) \textbf{Z}^{16}+(21913584 B_1 B_2^3 B_4^3 B_6^3 B_3^5+73045280 B_1^2 B_2^3 B_4^2 B_6^3 B_3^5+73045280 B_2^2 B_4^3 B_5^2 B_6^3 B_3^5+21913584
B_2^3 B_4^3 B_5 B_6^3 B_3^5+1016898960 B_1 B_2^2 B_4^3 B_5 B_6^3 B_3^5+1016898960 B_1 B_2^3 B_4^2 B_5 B_6^3 B_3^5+99607200 B_1^2 B_2^3 B_4^3 B_6^2
B_3^5+50820000 B_1^2 B_2^4 B_4^2 B_6^2 B_3^5+50820000 B_2^2 B_4^4 B_5^2 B_6^2 B_3^5+99607200 B_2^3 B_4^3 B_5^2 B_6^2 B_3^5+3252073440 B_1 B_2^2 B_4^3
B_5^2 B_6^2 B_3^5+2075150000 B_1 B_2^3 B_4^2 B_5^2 B_6^2 B_3^5+622545000 B_1 B_2^2 B_4^4 B_5 B_6^2 B_3^5+19480302576 B_1 B_2^3 B_4^3 B_5 B_6^2 B_3^5+2075150000
B_1^2 B_2^2 B_4^3 B_5 B_6^2 B_3^5+622545000 B_1 B_2^4 B_4^2 B_5 B_6^2 B_3^5+3252073440 B_1^2 B_2^3 B_4^2 B_5 B_6^2 B_3^5+887409600 B_1 B_2^3 B_4^3
B_5^2 B_6 B_3^5+887409600 B_1^2 B_2^3 B_4^3 B_5 B_6 B_3^5+871627680 B_1 B_2^2 B_4^3 B_5^2 B_6^3 B_3^4+766975440 B_1 B_2^3 B_4^2 B_5^2 B_6^3 B_3^4+2414513024
B_1 B_2^3 B_4^3 B_5 B_6^3 B_3^4+766975440 B_1^2 B_2^2 B_4^3 B_5 B_6^3 B_3^4+871627680 B_1^2 B_2^3 B_4^2 B_5 B_6^3 B_3^4+11383680 B_1 B_2^2 B_4^3
B_5^3 B_6^2 B_3^4+830060000 B_1 B_2^2 B_4^4 B_5^2 B_6^2 B_3^4+18476731056 B_1 B_2^3 B_4^3 B_5^2 B_6^2 B_3^4+4063973760 B_1^2 B_2^2 B_4^3 B_5^2 B_6^2
B_3^4+4063973760 B_1^2 B_2^3 B_4^2 B_5^2 B_6^2 B_3^4+1778700000 B_1 B_2^3 B_4^4 B_5 B_6^2 B_3^4+1778700000 B_1 B_2^4 B_4^3 B_5 B_6^2 B_3^4+18476731056
B_1^2 B_2^3 B_4^3 B_5 B_6^2 B_3^4+830060000 B_1^2 B_2^4 B_4^2 B_5 B_6^2 B_3^4+11383680 B_1^3 B_2^3 B_4^2 B_5 B_6^2 B_3^4+323400000 B_1 B_2^3 B_4^4
B_5^2 B_6 B_3^4+2253071744 B_1^2 B_2^3 B_4^3 B_5^2 B_6 B_3^4+323400000 B_1^2 B_2^4 B_4^3 B_5 B_6 B_3^4+4079910912 B_1^2 B_2^3 B_4^3 B_5^2 B_6^2 B_3^3)
\textbf{Z}^{15}+(41087970 B_1 B_2^3 B_4^2 B_6^3 B_3^5+41087970 B_2^2 B_4^3 B_5 B_6^3 B_3^5+136959900 B_1 B_2^2 B_4^2 B_5 B_6^3 B_3^5+56029050 B_1 B_2^3
B_4^3 B_6^2 B_3^5+186763500 B_1^2 B_2^3 B_4^2 B_6^2 B_3^5+186763500 B_2^2 B_4^3 B_5^2 B_6^2 B_3^5+56029050 B_2^3 B_4^3 B_5 B_6^2 B_3^5+2655776970
B_1 B_2^2 B_4^3 B_5 B_6^2 B_3^5+2655776970 B_1 B_2^3 B_4^2 B_5 B_6^2 B_3^5+499167900 B_1 B_2^3 B_4^3 B_5 B_6 B_3^5+35218260 B_1^2 B_2^3 B_4^2 B_6^3
B_3^4+35218260 B_2^2 B_4^3 B_5^2 B_6^3 B_3^4+117394200 B_1 B_2^2 B_4^2 B_5^2 B_6^3 B_3^4+607296690 B_1 B_2^2 B_4^3 B_5 B_6^3 B_3^4+607296690 B_1
B_2^3 B_4^2 B_5 B_6^3 B_3^4+117394200 B_1^2 B_2^2 B_4^2 B_5 B_6^3 B_3^4+48024900 B_1^2 B_2^3 B_4^3 B_6^2 B_3^4+24502500 B_1^2 B_2^4 B_4^2 B_6^2 B_3^4+24502500
B_2^2 B_4^4 B_5^2 B_6^2 B_3^4+48024900 B_2^3 B_4^3 B_5^2 B_6^2 B_3^4+4609356210 B_1 B_2^2 B_4^3 B_5^2 B_6^2 B_3^4+3054383640 B_1 B_2^3 B_4^2 B_5^2
B_6^2 B_3^4+446054400 B_1^2 B_2^2 B_4^2 B_5^2 B_6^2 B_3^4+300155625 B_1 B_2^2 B_4^4 B_5 B_6^2 B_3^4+14283282150 B_1 B_2^3 B_4^3 B_5 B_6^2 B_3^4+3054383640
B_1^2 B_2^2 B_4^3 B_5 B_6^2 B_3^4+300155625 B_1 B_2^4 B_4^2 B_5 B_6^2 B_3^4+4609356210 B_1^2 B_2^3 B_4^2 B_5 B_6^2 B_3^4+54573750 B_1 B_2^2 B_4^4
B_5^2 B_6 B_3^4+1671169500 B_1 B_2^3 B_4^3 B_5^2 B_6 B_3^4+298045440 B_1^2 B_2^2 B_4^3 B_5^2 B_6 B_3^4+298045440 B_1^2 B_2^3 B_4^2 B_5^2 B_6 B_3^4+1671169500
B_1^2 B_2^3 B_4^3 B_5 B_6 B_3^4+54573750 B_1^2 B_2^4 B_4^2 B_5 B_6 B_3^4+3585859200 B_1 B_2^3 B_4^3 B_5^2 B_6^2 B_3^3+1075757760 B_1^2 B_2^2 B_4^3
B_5^2 B_6^2 B_3^3+1075757760 B_1^2 B_2^3 B_4^2 B_5^2 B_6^2 B_3^3+3585859200 B_1^2 B_2^3 B_4^3 B_5 B_6^2 B_3^3+651974400 B_1^2 B_2^3 B_4^3 B_5^2 B_6
B_3^3) \textbf{Z}^{14}+(119528640 B_1 B_2^3 B_4^2 B_6^2 B_3^5+119528640 B_2^2 B_4^3 B_5 B_6^2 B_3^5+398428800 B_1 B_2^2 B_4^2 B_5 B_6^2 B_3^5+35218260
B_1 B_2^3 B_4^2 B_6^3 B_3^4+35218260 B_2^2 B_4^3 B_5 B_6^3 B_3^4+180457200 B_1 B_2^2 B_4^2 B_5 B_6^3 B_3^4+48024900 B_1 B_2^3 B_4^3 B_6^2 B_3^4+268939440
B_1^2 B_2^3 B_4^2 B_6^2 B_3^4+268939440 B_2^2 B_4^3 B_5^2 B_6^2 B_3^4+133402500 B_1 B_2 B_4^3 B_5^2 B_6^2 B_3^4+672348600 B_1 B_2^2 B_4^2 B_5^2 B_6^2
B_3^4+48024900 B_2^3 B_4^3 B_5 B_6^2 B_3^4+4320547560 B_1 B_2^2 B_4^3 B_5 B_6^2 B_3^4+4320547560 B_1 B_2^3 B_4^2 B_5 B_6^2 B_3^4+672348600 B_1^2
B_2^2 B_4^2 B_5 B_6^2 B_3^4+133402500 B_1^2 B_2^3 B_4 B_5 B_6^2 B_3^4+413887320 B_1 B_2^2 B_4^3 B_5^2 B_6 B_3^4+356548500 B_1 B_2^3 B_4^2 B_5^2 B_6
B_3^4+1189465200 B_1 B_2^3 B_4^3 B_5 B_6 B_3^4+356548500 B_1^2 B_2^2 B_4^3 B_5 B_6 B_3^4+413887320 B_1^2 B_2^3 B_4^2 B_5 B_6 B_3^4+1644128640 B_1
B_2^2 B_4^3 B_5^2 B_6^2 B_3^3+1075757760 B_1 B_2^3 B_4^2 B_5^2 B_6^2 B_3^3+292723200 B_1^2 B_2^2 B_4^2 B_5^2 B_6^2 B_3^3+3585859200 B_1 B_2^3 B_4^3
B_5 B_6^2 B_3^3+1075757760 B_1^2 B_2^2 B_4^3 B_5 B_6^2 B_3^3+1644128640 B_1^2 B_2^3 B_4^2 B_5 B_6^2 B_3^3+651974400 B_1 B_2^3 B_4^3 B_5^2 B_6 B_3^3+195592320
B_1^2 B_2^2 B_4^3 B_5^2 B_6 B_3^3+195592320 B_1^2 B_2^3 B_4^2 B_5^2 B_6 B_3^3+651974400 B_1^2 B_2^3 B_4^3 B_5 B_6 B_3^3) \textbf{Z}^{13}+(6391462
B_1 B_2^2 B_4^2 B_6^3 B_3^4+6391462 B_2^2 B_4^2 B_5 B_6^3 B_3^4+8715630 B_1 B_2^2 B_4^3 B_6^2 B_3^4+253998360 B_1 B_2^3 B_4^2 B_6^2 B_3^4+29052100
B_1^2 B_2^2 B_4^2 B_6^2 B_3^4+14822500 B_2 B_4^3 B_5^2 B_6^2 B_3^4+29052100 B_2^2 B_4^2 B_5^2 B_6^2 B_3^4+14822500 B_1^2 B_2^3 B_4 B_6^2 B_3^4+253998360
B_2^2 B_4^3 B_5 B_6^2 B_3^4+181575625 B_1 B_2 B_4^3 B_5 B_6^2 B_3^4+8715630 B_2^3 B_4^2 B_5 B_6^2 B_3^4+1554121926 B_1 B_2^2 B_4^2 B_5 B_6^2 B_3^4+181575625
B_1 B_2^3 B_4 B_5 B_6^2 B_3^4+20212500 B_1^2 B_2^3 B_4^2 B_6 B_3^4+20212500 B_2^2 B_4^3 B_5^2 B_6 B_3^4+110387200 B_1 B_2^2 B_4^2 B_5^2 B_6 B_3^4+388031490
B_1 B_2^2 B_4^3 B_5 B_6 B_3^4+388031490 B_1 B_2^3 B_4^2 B_5 B_6 B_3^4+110387200 B_1^2 B_2^2 B_4^2 B_5 B_6 B_3^4+5478396 B_1 B_2^2 B_4^2 B_5 B_6^3
B_3^3+111143340 B_1^2 B_2^3 B_4^2 B_6^2 B_3^3+111143340 B_2^2 B_4^3 B_5^2 B_6^2 B_3^3+155636250 B_1 B_2 B_4^3 B_5^2 B_6^2 B_3^3+542666124 B_1 B_2^2
B_4^2 B_5^2 B_6^2 B_3^3+1941993130 B_1 B_2^2 B_4^3 B_5 B_6^2 B_3^3+1941993130 B_1 B_2^3 B_4^2 B_5 B_6^2 B_3^3+542666124 B_1^2 B_2^2 B_4^2 B_5 B_6^2
B_3^3+155636250 B_1^2 B_2^3 B_4 B_5 B_6^2 B_3^3+353089660 B_1 B_2^2 B_4^3 B_5^2 B_6 B_3^3+247546530 B_1 B_2^3 B_4^2 B_5^2 B_6 B_3^3+60555264 B_1^2
B_2^2 B_4^2 B_5^2 B_6 B_3^3+707437500 B_1 B_2^3 B_4^3 B_5 B_6 B_3^3+247546530 B_1^2 B_2^2 B_4^3 B_5 B_6 B_3^3+353089660 B_1^2 B_2^3 B_4^2 B_5 B_6
B_3^3) \textbf{Z}^{12}+(66594528 B_1 B_2^2 B_4^2 B_6^2 B_3^4+21344400 B_1 B_2^3 B_4 B_6^2 B_3^4+21344400 B_2 B_4^3 B_5 B_6^2 B_3^4+66594528 B_2^2
B_4^2 B_5 B_6^2 B_3^4+71148000 B_1 B_2 B_4^2 B_5 B_6^2 B_3^4+71148000 B_1 B_2^2 B_4 B_5 B_6^2 B_3^4+29106000 B_1 B_2^3 B_4^2 B_6 B_3^4+29106000 B_2^2
B_4^3 B_5 B_6 B_3^4+191866752 B_1 B_2^2 B_4^2 B_5 B_6 B_3^4+137214000 B_1 B_2^3 B_4^2 B_6^2 B_3^3+35858592 B_1^2 B_2^2 B_4^2 B_6^2 B_3^3+18295200
B_2 B_4^3 B_5^2 B_6^2 B_3^3+35858592 B_2^2 B_4^2 B_5^2 B_6^2 B_3^3+60984000 B_1 B_2 B_4^2 B_5^2 B_6^2 B_3^3+18295200 B_1^2 B_2^3 B_4 B_6^2 B_3^3+137214000
B_2^2 B_4^3 B_5 B_6^2 B_3^3+224116200 B_1 B_2 B_4^3 B_5 B_6^2 B_3^3+1339753968 B_1 B_2^2 B_4^2 B_5 B_6^2 B_3^3+224116200 B_1 B_2^3 B_4 B_5 B_6^2
B_3^3+60984000 B_1^2 B_2^2 B_4 B_5 B_6^2 B_3^3+24948000 B_1^2 B_2^3 B_4^2 B_6 B_3^3+24948000 B_2^2 B_4^3 B_5^2 B_6 B_3^3+40748400 B_1 B_2 B_4^3 B_5^2
B_6 B_3^3+177031008 B_1 B_2^2 B_4^2 B_5^2 B_6 B_3^3+467082000 B_1 B_2^2 B_4^3 B_5 B_6 B_3^3+467082000 B_1 B_2^3 B_4^2 B_5 B_6 B_3^3+177031008 B_1^2
B_2^2 B_4^2 B_5 B_6 B_3^3+40748400 B_1^2 B_2^3 B_4 B_5 B_6 B_3^3) \textbf{Z}^{11}+(2614689 B_2^2 B_4^2 B_6^2 B_3^4+8715630 B_1 B_2^2 B_4 B_6^2
B_3^4+8715630 B_2 B_4^2 B_5 B_6^2 B_3^4+11884950 B_1 B_2^2 B_4^2 B_6 B_3^4+11884950 B_2^2 B_4^2 B_5 B_6 B_3^4+87268104 B_1 B_2^2 B_4^2 B_6^2 B_3^3+7470540
B_2 B_4^2 B_5^2 B_6^2 B_3^3+28586250 B_1 B_2^3 B_4 B_6^2 B_3^3+7470540 B_1^2 B_2^2 B_4 B_6^2 B_3^3+28586250 B_2 B_4^3 B_5 B_6^2 B_3^3+87268104 B_2^2
B_4^2 B_5 B_6^2 B_3^3+202848030 B_1 B_2 B_4^2 B_5 B_6^2 B_3^3+202848030 B_1 B_2^2 B_4 B_5 B_6^2 B_3^3+38981250 B_1 B_2^3 B_4^2 B_6 B_3^3+10187100
B_1^2 B_2^2 B_4^2 B_6 B_3^3+5197500 B_2 B_4^3 B_5^2 B_6 B_3^3+10187100 B_2^2 B_4^2 B_5^2 B_6 B_3^3+28385280 B_1 B_2 B_4^2 B_5^2 B_6 B_3^3+5197500
B_1^2 B_2^3 B_4 B_6 B_3^3+38981250 B_2^2 B_4^3 B_5 B_6 B_3^3+63669375 B_1 B_2 B_4^3 B_5 B_6 B_3^3+440527626 B_1 B_2^2 B_4^2 B_5 B_6 B_3^3+63669375
B_1 B_2^3 B_4 B_5 B_6 B_3^3+28385280 B_1^2 B_2^2 B_4 B_5 B_6 B_3^3+30735936 B_1 B_2^2 B_4^2 B_5 B_6^2 B_3^2+5588352 B_1 B_2^2 B_4^2 B_5^2 B_6 B_3^2+5588352
B_1^2 B_2^2 B_4^2 B_5 B_6 B_3^2) \textbf{Z}^{10}+(6225450 B_2^2 B_4^2 B_6^2 B_3^3+13280960 B_1 B_2 B_4^2 B_6^2 B_3^3+27544440 B_1 B_2^2 B_4 B_6^2
B_3^3+27544440 B_2 B_4^2 B_5 B_6^2 B_3^3+13280960 B_2^2 B_4 B_5 B_6^2 B_3^3+41164200 B_1 B_2 B_4 B_5 B_6^2 B_3^3+205800 B_1 B_2^2 B_4^2 B_5 B_3^3+37560600
B_1 B_2^2 B_4^2 B_6 B_3^3+3773000 B_2 B_4^2 B_5^2 B_6 B_3^3+9240000 B_1 B_2^3 B_4 B_6 B_3^3+3773000 B_1^2 B_2^2 B_4 B_6 B_3^3+9240000 B_2 B_4^3 B_5
B_6 B_3^3+37560600 B_2^2 B_4^2 B_5 B_6 B_3^3+82222140 B_1 B_2 B_4^2 B_5 B_6 B_3^3+82222140 B_1 B_2^2 B_4 B_5 B_6 B_3^3+11383680 B_1 B_2 B_4^2 B_5
B_6^2 B_3^2+11383680 B_1 B_2^2 B_4 B_5 B_6^2 B_3^2+2069760 B_1 B_2 B_4^2 B_5^2 B_6 B_3^2+24147200 B_1 B_2^2 B_4^2 B_5 B_6 B_3^2+2069760 B_1^2 B_2^2
B_4 B_5 B_6 B_3^2) \textbf{Z}^9+(1867635 B_2 B_4^2 B_6^2 B_3^3+1867635 B_2^2 B_4 B_6^2 B_3^3+6225450 B_1 B_2 B_4 B_6^2 B_3^3+6225450 B_2 B_4 B_5
B_6^2 B_3^3+2546775 B_2^2 B_4^2 B_6 B_3^3+8489250 B_1 B_2 B_4^2 B_6 B_3^3+13222440 B_1 B_2^2 B_4 B_6 B_3^3+13222440 B_2 B_4^2 B_5 B_6 B_3^3+8489250
B_2^2 B_4 B_5 B_6 B_3^3+23654400 B_1 B_2 B_4 B_5 B_6 B_3^3+1600830 B_1 B_2^2 B_4 B_6^2 B_3^2+1600830 B_2 B_4^2 B_5 B_6^2 B_3^2+5336100 B_1 B_2 B_4
B_5 B_6^2 B_3^2+396900 B_1 B_2^2 B_4^2 B_5 B_3^2+2182950 B_1 B_2^2 B_4^2 B_6 B_3^2+291060 B_2 B_4^2 B_5^2 B_6 B_3^2+291060 B_1^2 B_2^2 B_4 B_6 B_3^2+2182950
B_2^2 B_4^2 B_5 B_6 B_3^2+9168390 B_1 B_2 B_4^2 B_5 B_6 B_3^2+9168390 B_1 B_2^2 B_4 B_5 B_6 B_3^2) \textbf{Z}^8+(996072 B_2 B_4 B_6^2 B_3^3+1358280
B_2 B_4^2 B_6 B_3^3+1358280 B_2^2 B_4 B_6 B_3^3+4527600 B_1 B_2 B_4 B_6 B_3^3+4527600 B_2 B_4 B_5 B_6 B_3^3+853776 B_1 B_2 B_4 B_6^2 B_3^2+853776
B_2 B_4 B_5 B_6^2 B_3^2+211680 B_1 B_2 B_4^2 B_5 B_3^2+211680 B_1 B_2^2 B_4 B_5 B_3^2+1164240 B_1 B_2 B_4^2 B_6 B_3^2+1853280 B_1 B_2^2 B_4 B_6 B_3^2+1853280
B_2 B_4^2 B_5 B_6 B_3^2+1164240 B_2^2 B_4 B_5 B_6 B_3^2+6044544 B_1 B_2 B_4 B_5 B_6 B_3^2) \textbf{Z}^7+(916839 B_2 B_4 B_6 B_3^3+148225 B_2 B_4
B_6^2 B_3^2+36750 B_1 B_2^2 B_4 B_3^2+36750 B_2 B_4^2 B_5 B_3^2+200704 B_1 B_2 B_4 B_5 B_3^2+26950 B_1 B_2^2 B_6 B_3^2+202125 B_2 B_4^2 B_6 B_3^2+202125
B_2^2 B_4 B_6 B_3^2+1539384 B_1 B_2 B_4 B_6 B_3^2+26950 B_4^2 B_5 B_6 B_3^2+1539384 B_2 B_4 B_5 B_6 B_3^2+369600 B_1 B_2 B_4 B_5 B_6 B_3) \textbf{Z}^6+(44100
B_1 B_2 B_4 B_3^2+44100 B_2 B_4 B_5 B_3^2+32340 B_1 B_2 B_6 B_3^2+407484 B_2 B_4 B_6 B_3^2+32340 B_4 B_5 B_6 B_3^2+24192 B_1 B_2 B_4 B_5 B_3+133056
B_1 B_2 B_4 B_6 B_3+133056 B_2 B_4 B_5 B_6 B_3) \textbf{Z}^5+(11025 B_2 B_4 B_3^2+8085 B_2 B_6 B_3^2+8085 B_4 B_6 B_3^2+9450 B_1 B_2 B_4 B_3+9450
B_2 B_4 B_5 B_3+6930 B_1 B_2 B_6 B_3+51975 B_2 B_4 B_6 B_3+6930 B_4 B_5 B_6 B_3) \textbf{Z}^4+(560 B_1 B_2 B_3+4200 B_2 B_4 B_3+560 B_4 B_5 B_3+3080
B_2 B_6 B_3+3080 B_4 B_6 B_3) \textbf{Z}^3+(315 B_2 B_3+315 B_4 B_3+231 B_6 B_3) \textbf{Z}^2+42 B_3 \textbf{Z}+1\)

\vspace{1em}
\noindent\(\fbox{$H_4$}=B_1^3 B_2^6 B_3^8 B_4^6 B_5^3 B_6^4 \textbf{Z}^{30}+30 B_1^3 B_2^5 B_3^8 B_4^6 B_5^3 B_6^4 \textbf{Z}^{29}+(120 B_1^2 B_2^5 B_4^6 B_5^3
B_6^4 B_3^8+315 B_1^3 B_2^5 B_4^6 B_5^3 B_6^4 B_3^7) \textbf{Z}^{28}+(2240 B_1^2 B_2^5 B_4^6 B_5^3 B_6^4 B_3^7+1050 B_1^3 B_2^5 B_4^5 B_5^3 B_6^4
B_3^7+770 B_1^3 B_2^5 B_4^6 B_5^3 B_6^3 B_3^7) \textbf{Z}^{27}+(4200 B_1^2 B_2^4 B_4^6 B_5^3 B_6^4 B_3^7+9450 B_1^2 B_2^5 B_4^5 B_5^3 B_6^4 B_3^7+1050
B_1^3 B_2^5 B_4^5 B_5^2 B_6^4 B_3^7+6930 B_1^2 B_2^5 B_4^6 B_5^3 B_6^3 B_3^7+5775 B_1^3 B_2^5 B_4^5 B_5^3 B_6^3 B_3^7) \textbf{Z}^{26}+(31500 B_1^2
B_2^4 B_4^5 B_5^3 B_6^4 B_3^7+10752 B_1^2 B_2^5 B_4^5 B_5^2 B_6^4 B_3^7+23100 B_1^2 B_2^4 B_4^6 B_5^3 B_6^3 B_3^7+59136 B_1^2 B_2^5 B_4^5 B_5^3 B_6^3
B_3^7+8316 B_1^3 B_2^5 B_4^5 B_5^2 B_6^3 B_3^7+9702 B_1^3 B_2^5 B_4^5 B_5^3 B_6^3 B_3^6) \textbf{Z}^{25}+(45360 B_1^2 B_2^4 B_4^5 B_5^2 B_6^4 B_3^7+249480
B_1^2 B_2^4 B_4^5 B_5^3 B_6^3 B_3^7+92400 B_1^2 B_2^5 B_4^5 B_5^2 B_6^3 B_3^7+36750 B_1^2 B_2^4 B_4^5 B_5^3 B_6^4 B_3^6+26950 B_1^2 B_2^4 B_4^6 B_5^3
B_6^3 B_3^6+107800 B_1^2 B_2^5 B_4^5 B_5^3 B_6^3 B_3^6+8085 B_1^3 B_2^4 B_4^5 B_5^3 B_6^3 B_3^6+26950 B_1^3 B_2^5 B_4^5 B_5^2 B_6^3 B_3^6)
\textbf{Z}^{24}+(443520 B_1^2 B_2^4 B_4^5 B_5^2 B_6^3 B_3^7+94080 B_1^2 B_2^4 B_4^5 B_5^2 B_6^4 B_3^6+1132560 B_1^2 B_2^4 B_4^5 B_5^3 B_6^3 B_3^6+316800
B_1^2 B_2^5 B_4^5 B_5^2 B_6^3 B_3^6+32340 B_1^3 B_2^4 B_4^5 B_5^2 B_6^3 B_3^6+16500 B_1^3 B_2^5 B_4^4 B_5^2 B_6^3 B_3^6) \textbf{Z}^{23}+(44100
B_1^2 B_2^4 B_4^4 B_5^2 B_6^4 B_3^6+32340 B_1 B_2^4 B_4^5 B_5^3 B_6^3 B_3^6+495000 B_1^2 B_2^3 B_4^5 B_5^3 B_6^3 B_3^6+242550 B_1^2 B_2^4 B_4^4 B_5^3
B_6^3 B_3^6+3256110 B_1^2 B_2^4 B_4^5 B_5^2 B_6^3 B_3^6+202125 B_1^2 B_2^5 B_4^4 B_5^2 B_6^3 B_3^6+44550 B_1^3 B_2^4 B_4^4 B_5^2 B_6^3 B_3^6+177870
B_1^2 B_2^4 B_4^5 B_5^3 B_6^2 B_3^6+1358280 B_1^2 B_2^4 B_4^5 B_5^3 B_6^3 B_3^5) \textbf{Z}^{22}+(178200 B_1 B_2^3 B_4^5 B_5^3 B_6^3 B_3^6+168960
B_1 B_2^4 B_4^5 B_5^2 B_6^3 B_3^6+2182950 B_1^2 B_2^3 B_4^5 B_5^2 B_6^3 B_3^6+2674100 B_1^2 B_2^4 B_4^4 B_5^2 B_6^3 B_3^6+711480 B_1^2 B_2^4 B_4^5
B_5^2 B_6^2 B_3^6+1478400 B_1^2 B_2^3 B_4^5 B_5^3 B_6^3 B_3^5+1131900 B_1^2 B_2^4 B_4^4 B_5^3 B_6^3 B_3^5+4928000 B_1^2 B_2^4 B_4^5 B_5^2 B_6^3 B_3^5+23100
B_1^3 B_2^4 B_4^4 B_5^2 B_6^3 B_3^5+830060 B_1^2 B_2^4 B_4^5 B_5^3 B_6^2 B_3^5) \textbf{Z}^{21}+(970200 B_1 B_2^3 B_4^5 B_5^2 B_6^3 B_3^6+349272
B_1 B_2^4 B_4^4 B_5^2 B_6^3 B_3^6+3234000 B_1^2 B_2^3 B_4^4 B_5^2 B_6^3 B_3^6+155232 B_1^2 B_2^4 B_4^4 B_5 B_6^3 B_3^6+853776 B_1^2 B_2^4 B_4^4 B_5^2
B_6^2 B_3^6+577500 B_1 B_2^3 B_4^5 B_5^3 B_6^3 B_3^5+1559250 B_1^2 B_2^3 B_4^4 B_5^3 B_6^3 B_3^5+7074375 B_1^2 B_2^3 B_4^5 B_5^2 B_6^3 B_3^5+9315306
B_1^2 B_2^4 B_4^4 B_5^2 B_6^3 B_3^5+1143450 B_1^2 B_2^3 B_4^5 B_5^3 B_6^2 B_3^5+996072 B_1^2 B_2^4 B_4^4 B_5^3 B_6^2 B_3^5+3811500 B_1^2 B_2^4 B_4^5
B_5^2 B_6^2 B_3^5+5082 B_1^3 B_2^4 B_4^4 B_5^2 B_6^2 B_3^5) \textbf{Z}^{20}+(2069760 B_1 B_2^3 B_4^4 B_5^2 B_6^3 B_3^6+693000 B_1 B_2^3 B_4^4 B_5^3
B_6^3 B_3^5+3326400 B_1 B_2^3 B_4^5 B_5^2 B_6^3 B_3^5+369600 B_1 B_2^4 B_4^4 B_5^2 B_6^3 B_3^5+21801780 B_1^2 B_2^3 B_4^4 B_5^2 B_6^3 B_3^5+4331250
B_1^2 B_2^4 B_4^3 B_5^2 B_6^3 B_3^5+1478400 B_1^2 B_2^4 B_4^4 B_5 B_6^3 B_3^5+508200 B_1 B_2^3 B_4^5 B_5^3 B_6^2 B_3^5+2439360 B_1^2 B_2^3 B_4^4
B_5^3 B_6^2 B_3^5+6225450 B_1^2 B_2^3 B_4^5 B_5^2 B_6^2 B_3^5+11384100 B_1^2 B_2^4 B_4^4 B_5^2 B_6^2 B_3^5) \textbf{Z}^{19}+(14314300 B_1 B_2^3
B_4^4 B_5^2 B_6^3 B_3^5+14437500 B_1^2 B_2^3 B_4^3 B_5^2 B_6^3 B_3^5+3056130 B_1^2 B_2^3 B_4^4 B_5 B_6^3 B_3^5+1559250 B_1^2 B_2^4 B_4^3 B_5 B_6^3
B_3^5+1372140 B_1 B_2^3 B_4^4 B_5^3 B_6^2 B_3^5+3176250 B_1 B_2^3 B_4^5 B_5^2 B_6^2 B_3^5+127050 B_1 B_2^4 B_4^4 B_5^2 B_6^2 B_3^5+28420210 B_1^2
B_2^3 B_4^4 B_5^2 B_6^2 B_3^5+8575875 B_1^2 B_2^4 B_4^3 B_5^2 B_6^2 B_3^5+2032800 B_1^2 B_2^4 B_4^4 B_5 B_6^2 B_3^5+6338640 B_1^2 B_2^3 B_4^4 B_5^2
B_6^3 B_3^4+711480 B_1^2 B_2^3 B_4^4 B_5^3 B_6^2 B_3^4+2371600 B_1^2 B_2^4 B_4^4 B_5^2 B_6^2 B_3^4) \textbf{Z}^{18}+(577500 B_1 B_2^2 B_4^4 B_5^2
B_6^3 B_3^5+10187100 B_1 B_2^3 B_4^3 B_5^2 B_6^3 B_3^5+1774080 B_1 B_2^3 B_4^4 B_5 B_6^3 B_3^5+5913600 B_1^2 B_2^3 B_4^3 B_5 B_6^3 B_3^5+18705960
B_1 B_2^3 B_4^4 B_5^2 B_6^2 B_3^5+32524800 B_1^2 B_2^3 B_4^3 B_5^2 B_6^2 B_3^5+7470540 B_1^2 B_2^3 B_4^4 B_5 B_6^2 B_3^5+3811500 B_1^2 B_2^4 B_4^3
B_5 B_6^2 B_3^5+8279040 B_1 B_2^3 B_4^4 B_5^2 B_6^3 B_3^4+8731800 B_1^2 B_2^3 B_4^3 B_5^2 B_6^3 B_3^4+711480 B_1 B_2^3 B_4^4 B_5^3 B_6^2 B_3^4+16625700
B_1^2 B_2^3 B_4^4 B_5^2 B_6^2 B_3^4+4446750 B_1^2 B_2^4 B_4^3 B_5^2 B_6^2 B_3^4) \textbf{Z}^{17}+(4527600 B_1 B_2^3 B_4^3 B_5 B_6^3 B_3^5+508200
B_1 B_2^2 B_4^4 B_5^2 B_6^2 B_3^5+24901800 B_1 B_2^3 B_4^3 B_5^2 B_6^2 B_3^5+5488560 B_1 B_2^3 B_4^4 B_5 B_6^2 B_3^5+18295200 B_1^2 B_2^3 B_4^3 B_5
B_6^2 B_3^5+2182950 B_1 B_2^2 B_4^4 B_5^2 B_6^3 B_3^4+11884950 B_1 B_2^3 B_4^3 B_5^2 B_6^3 B_3^4+3880800 B_1^2 B_2^3 B_4^3 B_5 B_6^3 B_3^4+108900
B_1 B_2^2 B_4^4 B_5^3 B_6^2 B_3^4+18478980 B_1 B_2^3 B_4^4 B_5^2 B_6^2 B_3^4+1334025 B_1^2 B_2^2 B_4^4 B_5^2 B_6^2 B_3^4+45530550 B_1^2 B_2^3 B_4^3
B_5^2 B_6^2 B_3^4+4446750 B_1^2 B_2^3 B_4^4 B_5 B_6^2 B_3^4+2268750 B_1^2 B_2^4 B_4^3 B_5 B_6^2 B_3^4+1584660 B_1^2 B_2^3 B_4^4 B_5^2 B_6 B_3^4)
\textbf{Z}^{16}+(15937152 B_1 B_2^3 B_4^3 B_5 B_6^2 B_3^5+3234000 B_1 B_2^2 B_4^3 B_5^2 B_6^3 B_3^4+5588352 B_1 B_2^3 B_4^3 B_5 B_6^3 B_3^4+6203600 B_1
B_2^2 B_4^4 B_5^2 B_6^2 B_3^4+53742416 B_1 B_2^3 B_4^3 B_5^2 B_6^2 B_3^4+5808000 B_1^2 B_2^2 B_4^3 B_5^2 B_6^2 B_3^4+5808000 B_1 B_2^3 B_4^4 B_5
B_6^2 B_3^4+34036496 B_1^2 B_2^3 B_4^3 B_5 B_6^2 B_3^4+3234000 B_1 B_2^3 B_4^4 B_5^2 B_6 B_3^4+5588352 B_1^2 B_2^3 B_4^3 B_5^2 B_6 B_3^4+15937152
B_1^2 B_2^3 B_4^3 B_5^2 B_6^2 B_3^3) \textbf{Z}^{15}+(1584660 B_1 B_2^2 B_4^3 B_5 B_6^3 B_3^4+108900 B_2^2 B_4^4 B_5^2 B_6^2 B_3^4+18478980 B_1
B_2^2 B_4^3 B_5^2 B_6^2 B_3^4+1334025 B_1 B_2^2 B_4^4 B_5 B_6^2 B_3^4+45530550 B_1 B_2^3 B_4^3 B_5 B_6^2 B_3^4+4446750 B_1^2 B_2^2 B_4^3 B_5 B_6^2
B_3^4+2268750 B_1^2 B_2^3 B_4^2 B_5 B_6^2 B_3^4+2182950 B_1 B_2^2 B_4^4 B_5^2 B_6 B_3^4+11884950 B_1 B_2^3 B_4^3 B_5^2 B_6 B_3^4+3880800 B_1^2 B_2^3
B_4^3 B_5 B_6 B_3^4+508200 B_1 B_2^2 B_4^4 B_5^2 B_6^2 B_3^3+24901800 B_1 B_2^3 B_4^3 B_5^2 B_6^2 B_3^3+5488560 B_1^2 B_2^2 B_4^3 B_5^2 B_6^2 B_3^3+18295200
B_1^2 B_2^3 B_4^3 B_5 B_6^2 B_3^3+4527600 B_1^2 B_2^3 B_4^3 B_5^2 B_6 B_3^3) \textbf{Z}^{14}+(711480 B_2^2 B_4^3 B_5^2 B_6^2 B_3^4+16625700 B_1
B_2^2 B_4^3 B_5 B_6^2 B_3^4+4446750 B_1 B_2^3 B_4^2 B_5 B_6^2 B_3^4+8279040 B_1 B_2^2 B_4^3 B_5^2 B_6 B_3^4+8731800 B_1 B_2^3 B_4^3 B_5 B_6 B_3^4+18705960
B_1 B_2^2 B_4^3 B_5^2 B_6^2 B_3^3+32524800 B_1 B_2^3 B_4^3 B_5 B_6^2 B_3^3+7470540 B_1^2 B_2^2 B_4^3 B_5 B_6^2 B_3^3+3811500 B_1^2 B_2^3 B_4^2 B_5
B_6^2 B_3^3+577500 B_1 B_2^2 B_4^4 B_5^2 B_6 B_3^3+10187100 B_1 B_2^3 B_4^3 B_5^2 B_6 B_3^3+1774080 B_1^2 B_2^2 B_4^3 B_5^2 B_6 B_3^3+5913600 B_1^2
B_2^3 B_4^3 B_5 B_6 B_3^3) \textbf{Z}^{13}+(711480 B_2^2 B_4^3 B_5 B_6^2 B_3^4+2371600 B_1 B_2^2 B_4^2 B_5 B_6^2 B_3^4+6338640 B_1 B_2^2 B_4^3
B_5 B_6 B_3^4+1372140 B_2^2 B_4^3 B_5^2 B_6^2 B_3^3+3176250 B_1 B_2 B_4^3 B_5^2 B_6^2 B_3^3+127050 B_1 B_2^2 B_4^2 B_5^2 B_6^2 B_3^3+28420210 B_1
B_2^2 B_4^3 B_5 B_6^2 B_3^3+8575875 B_1 B_2^3 B_4^2 B_5 B_6^2 B_3^3+2032800 B_1^2 B_2^2 B_4^2 B_5 B_6^2 B_3^3+14314300 B_1 B_2^2 B_4^3 B_5^2 B_6
B_3^3+14437500 B_1 B_2^3 B_4^3 B_5 B_6 B_3^3+3056130 B_1^2 B_2^2 B_4^3 B_5 B_6 B_3^3+1559250 B_1^2 B_2^3 B_4^2 B_5 B_6 B_3^3) \textbf{Z}^{12}+(508200
B_2 B_4^3 B_5^2 B_6^2 B_3^3+2439360 B_2^2 B_4^3 B_5 B_6^2 B_3^3+6225450 B_1 B_2 B_4^3 B_5 B_6^2 B_3^3+11384100 B_1 B_2^2 B_4^2 B_5 B_6^2 B_3^3+693000
B_2^2 B_4^3 B_5^2 B_6 B_3^3+3326400 B_1 B_2 B_4^3 B_5^2 B_6 B_3^3+369600 B_1 B_2^2 B_4^2 B_5^2 B_6 B_3^3+21801780 B_1 B_2^2 B_4^3 B_5 B_6 B_3^3+4331250
B_1 B_2^3 B_4^2 B_5 B_6 B_3^3+1478400 B_1^2 B_2^2 B_4^2 B_5 B_6 B_3^3+2069760 B_1 B_2^2 B_4^3 B_5^2 B_6 B_3^2) \textbf{Z}^{11}+(5082 B_1 B_2^2
B_4^2 B_6^2 B_3^3+1143450 B_2 B_4^3 B_5 B_6^2 B_3^3+996072 B_2^2 B_4^2 B_5 B_6^2 B_3^3+3811500 B_1 B_2 B_4^2 B_5 B_6^2 B_3^3+577500 B_2 B_4^3 B_5^2
B_6 B_3^3+1559250 B_2^2 B_4^3 B_5 B_6 B_3^3+7074375 B_1 B_2 B_4^3 B_5 B_6 B_3^3+9315306 B_1 B_2^2 B_4^2 B_5 B_6 B_3^3+853776 B_1 B_2^2 B_4^2 B_5
B_6^2 B_3^2+970200 B_1 B_2 B_4^3 B_5^2 B_6 B_3^2+349272 B_1 B_2^2 B_4^2 B_5^2 B_6 B_3^2+3234000 B_1 B_2^2 B_4^3 B_5 B_6 B_3^2+155232 B_1^2 B_2^2
B_4^2 B_5 B_6 B_3^2) \textbf{Z}^{10}+(830060 B_2 B_4^2 B_5 B_6^2 B_3^3+23100 B_1 B_2^2 B_4^2 B_6 B_3^3+1478400 B_2 B_4^3 B_5 B_6 B_3^3+1131900
B_2^2 B_4^2 B_5 B_6 B_3^3+4928000 B_1 B_2 B_4^2 B_5 B_6 B_3^3+711480 B_1 B_2 B_4^2 B_5 B_6^2 B_3^2+178200 B_2 B_4^3 B_5^2 B_6 B_3^2+168960 B_1 B_2
B_4^2 B_5^2 B_6 B_3^2+2182950 B_1 B_2 B_4^3 B_5 B_6 B_3^2+2674100 B_1 B_2^2 B_4^2 B_5 B_6 B_3^2) \textbf{Z}^9+(1358280 B_2 B_4^2 B_5 B_6 B_3^3+177870
B_2 B_4^2 B_5 B_6^2 B_3^2+44100 B_1 B_2^2 B_4^2 B_5 B_3^2+44550 B_1 B_2^2 B_4^2 B_6 B_3^2+32340 B_2 B_4^2 B_5^2 B_6 B_3^2+495000 B_2 B_4^3 B_5 B_6
B_3^2+242550 B_2^2 B_4^2 B_5 B_6 B_3^2+3256110 B_1 B_2 B_4^2 B_5 B_6 B_3^2+202125 B_1 B_2^2 B_4 B_5 B_6 B_3^2) \textbf{Z}^8+(94080 B_1 B_2 B_4^2
B_5 B_3^2+32340 B_1 B_2 B_4^2 B_6 B_3^2+16500 B_1 B_2^2 B_4 B_6 B_3^2+1132560 B_2 B_4^2 B_5 B_6 B_3^2+316800 B_1 B_2 B_4 B_5 B_6 B_3^2+443520 B_1
B_2 B_4^2 B_5 B_6 B_3) \textbf{Z}^7+(36750 B_2 B_4^2 B_5 B_3^2+8085 B_2 B_4^2 B_6 B_3^2+26950 B_1 B_2 B_4 B_6 B_3^2+26950 B_4^2 B_5 B_6 B_3^2+107800
B_2 B_4 B_5 B_6 B_3^2+45360 B_1 B_2 B_4^2 B_5 B_3+249480 B_2 B_4^2 B_5 B_6 B_3+92400 B_1 B_2 B_4 B_5 B_6 B_3) \textbf{Z}^6+(9702 B_2 B_4 B_6 B_3^2+31500
B_2 B_4^2 B_5 B_3+10752 B_1 B_2 B_4 B_5 B_3+8316 B_1 B_2 B_4 B_6 B_3+23100 B_4^2 B_5 B_6 B_3+59136 B_2 B_4 B_5 B_6 B_3) \textbf{Z}^5+(4200 B_3
B_5 B_4^2+1050 B_1 B_2 B_3 B_4+9450 B_2 B_3 B_5 B_4+5775 B_2 B_3 B_6 B_4+6930 B_3 B_5 B_6 B_4) \textbf{Z}^4+(1050 B_2 B_3 B_4+2240 B_3 B_5 B_4+770
B_3 B_6 B_4) \textbf{Z}^3+(315 B_3 B_4+120 B_5 B_4) \textbf{Z}^2+30 B_4 \textbf{Z}+1\)

\vspace{1em}
\noindent\(\fbox{$H_5$}=B_1^2 B_2^3 B_3^4 B_4^3 B_5^2 B_6^2 \textbf{Z}^{16}+16 B_1 B_2^3 B_3^4 B_4^3 B_5^2 B_6^2 \textbf{Z}^{15}+120 B_1 B_2^2 B_3^4 B_4^3 B_5^2 B_6^2
\textbf{Z}^{14}+560 B_1 B_2^2 B_3^3 B_4^3 B_5^2 B_6^2 \textbf{Z}^{13}+(1050 B_1 B_2^2 B_4^2 B_5^2 B_6^2 B_3^3+770 B_1 B_2^2 B_4^3 B_5^2 B_6 B_3^3) \textbf{Z}^{12}+(672
B_1 B_2^2 B_4^2 B_5 B_6^2 B_3^3+3696 B_1 B_2^2 B_4^2 B_5^2 B_6 B_3^3) \textbf{Z}^{11}+(3696 B_1 B_2^2 B_4^2 B_5 B_6 B_3^3+4312 B_1 B_2^2 B_4^2
B_5^2 B_6 B_3^2) \textbf{Z}^{10}+(2640 B_1 B_2 B_3^2 B_5^2 B_6 B_4^2+8800 B_1 B_2^2 B_3^2 B_5 B_6 B_4^2) \textbf{Z}^9+(660 B_2 B_4^2 B_5^2 B_6
B_3^2+8085 B_1 B_2 B_4^2 B_5 B_6 B_3^2+4125 B_1 B_2^2 B_4 B_5 B_6 B_3^2) \textbf{Z}^8+(2640 B_2 B_4^2 B_5 B_6 B_3^2+8800 B_1 B_2 B_4 B_5 B_6 B_3^2)
\textbf{Z}^7+(4312 B_2 B_4 B_5 B_6 B_3^2+3696 B_1 B_2 B_4 B_5 B_6 B_3) \textbf{Z}^6+(672 B_1 B_2 B_3 B_4 B_5+3696 B_2 B_3 B_4 B_6 B_5) \textbf{Z}^5+(1050
B_2 B_3 B_4 B_5+770 B_3 B_4 B_6 B_5) \textbf{Z}^4+560 B_3 B_4 B_5 \textbf{Z}^3+120 B_4 B_5 \textbf{Z}^2+16 B_5 \textbf{Z}+1\)

\vspace{1em}
\noindent\(\fbox{$H_6$}=B_1^2 B_2^4 B_3^6 B_4^4 B_5^2 B_6^4 \textbf{Z}^{22}+22 B_1^2 B_2^4 B_3^6 B_4^4 B_5^2 B_6^3 \textbf{Z}^{21}+231 B_1^2 B_2^4 B_3^5 B_4^4 B_5^2
B_6^3 \textbf{Z}^{20}+(770 B_1^2 B_2^3 B_4^4 B_5^2 B_6^3 B_3^5+770 B_1^2 B_2^4 B_4^3 B_5^2 B_6^3 B_3^5) \textbf{Z}^{19}+(770 B_1 B_2^3 B_4^4 B_5^2
B_6^3 B_3^5+5775 B_1^2 B_2^3 B_4^3 B_5^2 B_6^3 B_3^5+770 B_1^2 B_2^4 B_4^3 B_5 B_6^3 B_3^5) \textbf{Z}^{18}+(8316 B_1 B_2^3 B_4^3 B_5^2 B_6^3 B_3^5+8316
B_1^2 B_2^3 B_4^3 B_5 B_6^3 B_3^5+9702 B_1^2 B_2^3 B_4^3 B_5^2 B_6^3 B_3^4) \textbf{Z}^{17}+(14784 B_1 B_2^3 B_4^3 B_5 B_6^3 B_3^5+26950 B_1 B_2^3
B_4^3 B_5^2 B_6^3 B_3^4+26950 B_1^2 B_2^3 B_4^3 B_5 B_6^3 B_3^4+5929 B_1^2 B_2^3 B_4^3 B_5^2 B_6^2 B_3^4) \textbf{Z}^{16}+(16500 B_1 B_2^2 B_4^3
B_5^2 B_6^3 B_3^4+90112 B_1 B_2^3 B_4^3 B_5 B_6^3 B_3^4+16500 B_1^2 B_2^3 B_4^2 B_5 B_6^3 B_3^4+23716 B_1 B_2^3 B_4^3 B_5^2 B_6^2 B_3^4+23716 B_1^2
B_2^3 B_4^3 B_5 B_6^2 B_3^4) \textbf{Z}^{15}+(72765 B_1 B_2^2 B_4^3 B_5 B_6^3 B_3^4+72765 B_1 B_2^3 B_4^2 B_5 B_6^3 B_3^4+32670 B_1 B_2^2 B_4^3
B_5^2 B_6^2 B_3^4+108900 B_1 B_2^3 B_4^3 B_5 B_6^2 B_3^4+32670 B_1^2 B_2^3 B_4^2 B_5 B_6^2 B_3^4) \textbf{Z}^{14}+(107800 B_1 B_2^2 B_4^2 B_5 B_6^3
B_3^4+177870 B_1 B_2^2 B_4^3 B_5 B_6^2 B_3^4+177870 B_1 B_2^3 B_4^2 B_5 B_6^2 B_3^4+16940 B_1 B_2^2 B_4^3 B_5^2 B_6^2 B_3^3+16940 B_1^2 B_2^3 B_4^2
B_5 B_6^2 B_3^3) \textbf{Z}^{13}+(379456 B_1 B_2^2 B_4^2 B_5 B_6^2 B_3^4+45276 B_1 B_2^2 B_4^2 B_5 B_6^3 B_3^3+5082 B_1 B_2^2 B_4^2 B_5^2 B_6^2
B_3^3+105875 B_1 B_2^2 B_4^3 B_5 B_6^2 B_3^3+105875 B_1 B_2^3 B_4^2 B_5 B_6^2 B_3^3+5082 B_1^2 B_2^2 B_4^2 B_5 B_6^2 B_3^3) \textbf{Z}^{12}+705432 B_1
B_2^2 B_3^3 B_4^2 B_5 B_6^2 \textbf{Z}^{11}+(5082 B_1 B_2^2 B_4^2 B_6^2 B_3^3+5082 B_2^2 B_4^2 B_5 B_6^2 B_3^3+105875 B_1 B_2 B_4^2 B_5 B_6^2 B_3^3+105875
B_1 B_2^2 B_4 B_5 B_6^2 B_3^3+45276 B_1 B_2^2 B_4^2 B_5 B_6 B_3^3+379456 B_1 B_2^2 B_4^2 B_5 B_6^2 B_3^2) \textbf{Z}^{10}+(16940 B_1 B_2^2 B_4
B_6^2 B_3^3+16940 B_2 B_4^2 B_5 B_6^2 B_3^3+177870 B_1 B_2 B_4^2 B_5 B_6^2 B_3^2+177870 B_1 B_2^2 B_4 B_5 B_6^2 B_3^2+107800 B_1 B_2^2 B_4^2 B_5
B_6 B_3^2) \textbf{Z}^9+(32670 B_1 B_2^2 B_4 B_6^2 B_3^2+32670 B_2 B_4^2 B_5 B_6^2 B_3^2+108900 B_1 B_2 B_4 B_5 B_6^2 B_3^2+72765 B_1 B_2 B_4^2
B_5 B_6 B_3^2+72765 B_1 B_2^2 B_4 B_5 B_6 B_3^2) \textbf{Z}^8+(23716 B_1 B_2 B_4 B_6^2 B_3^2+23716 B_2 B_4 B_5 B_6^2 B_3^2+16500 B_1 B_2^2 B_4
B_6 B_3^2+16500 B_2 B_4^2 B_5 B_6 B_3^2+90112 B_1 B_2 B_4 B_5 B_6 B_3^2) \textbf{Z}^7+(5929 B_2 B_4 B_6^2 B_3^2+26950 B_1 B_2 B_4 B_6 B_3^2+26950
B_2 B_4 B_5 B_6 B_3^2+14784 B_1 B_2 B_4 B_5 B_6 B_3) \textbf{Z}^6+(9702 B_2 B_4 B_6 B_3^2+8316 B_1 B_2 B_4 B_6 B_3+8316 B_2 B_4 B_5 B_6 B_3)
\textbf{Z}^5+(770 B_1 B_2 B_3 B_6+5775 B_2 B_3 B_4 B_6+770 B_3 B_4 B_5 B_6) \textbf{Z}^4+(770 B_2 B_3 B_6+770 B_3 B_4 B_6) \textbf{Z}^3+231 B_3 B_6 \textbf{Z}^2+22
B_6 \textbf{Z}+1\)

It should be noted that five polynomials: $H_1, H_2, H_4, H_5, H_6$ were found earlier (in a non-ordered form)
in ref. \cite{Top}. The biggest key polynomial $H_3$ was not presented in ref. \cite{Top}. 
We note that the ``length'' of the polynomial $H_3$  is more than $5/8$ of the total ``length'' of all polynomials.

%{\large

 \end{document}